%% file: ms.tex
\documentclass[useAmain sequence,usenatbib]{mn2e}

\usepackage[colorlinks=true,citecolor=blue]{hyperref}
 \usepackage{graphicx}

\usepackage{subfigure}
\usepackage{float}
\usepackage{placeins}
\usepackage{multicol}
\usepackage{multirow}
\usepackage{bm}
\usepackage{longtable}
\usepackage{rotating}
\usepackage{pdflscape}

\DeclareMathAlphabet{\mathsc}{OT1}{cmr}{m}{sc}
\def\testbx{bx}
\DeclareRobustCommand{\ion}[2]{
\relax\ifmmode
\ifx\testbx\f@series
{\mathbf{#1\,\mathsc{#2}}}\else
{\mathrm{#1\,\mathsc{#2}}}\fi
\else\textup{#1\,{\mdseries\textsc{#2}}}
\fi}
\newcommand{\Hi} {\ion{H}{i}}

\newcommand{\ha} {\mbox{H$\alpha$}}
\newcommand{\hb} {\mbox{H$\beta$}}

\newcommand{\Heii} {\ion{He}{ii}}

\newcommand{\Nai} {\ion{Na}{i}}

\def\snhj{SN 2013hj}
\def\sng{SN 2014G}
\def\hosthj{MCG -02-24-3}
\def\hostg{NGC 3448}
\def\thj{2456637.0}
\def\tg{2456669.7}
\def\sne{SNe 2013hj and 2014G}

\newcommand{\daophot}{\textsc{daophot}}

\newcommand{\epm}{\textsc{epm}}
\newcommand{\ubvri}{\textit{UBVRI}}
\newcommand{\swift}{\textit{Swift}}

\newcommand{\ebv}{\mbox{$E(B-V)$}}

\newcommand{\degree}{\mbox{$^\circ$}}
\newcommand{\maghundred}{\mbox{mag (100 d)$ ^{-1} $}}
\newcommand{\msun}{\mbox{M$_{\odot}$}}

\newcommand{\rsun}{\mbox{R$_{\odot}$}}
\newcommand{\kms}{\mbox{$\rm{\,km\,s^{-1}}$}}

\newcommand{\nickel}{\mbox{$^{56}$Ni}}
\newcommand{\cobalt}{\mbox{$^{56}$Co}}
\newcommand{\iron}{\mbox{$^{56}$Fe}}

\newcommand{\el}{\mbox{${e}^{-}$}}

\begin{document}

\title[Type II \sne]
{Photometric and polarimetric observations of fast declining Type II supernovae 2013hj and 2014G}
\author[Bose et al.]
{Subhash Bose$^{1,2}$\thanks{e-mail: email@subhashbose.com, bose@aries.res.in},
Brijesh Kumar$^1$, Kuntal Misra$^1$, Katsura Matsumoto$^3$\thanks{Author's contribution is for \snhj.},
\newauthor
Brajesh Kumar$^{4,1}$, Mridweeka Singh$ ^1 $,
Daiki Fukushima$^3$\footnotemark[2], Miho Kawabata$^3$\footnotemark[2]
\\
 $^1$Aryabhatta Research Institute of Observational Sciences, Manora
    Peak, Nainital - 263002, India.\\
 $^2$Centre of Advance Study, Department of Physics, Kumaun University, Nainital - 263001, India.\\
 $^3$Osaka Kyoiku University, 4-698-1 Asahigaoka, Kashiwara, Osaka 582-8582\\
 $^4$Indian Institute of Astrophysics, Block-II, Koramangala, Bangalore - 560034, India.\\
}

\date{Accepted.....; Received .....}

\pagerange{\pageref{firstpage}--\pageref{lastpage}} \pubyear{}

\maketitle

\label{firstpage}

\begin{abstract}
We present broadband photometric and polarimetric observations of two type II supernovae (SNe) 2013hj and 2014G. \sng\ is a spectroscopically classified type IIL event, which we also confirm photometrically as its light curve show characteristic features (plateau slope of 2.55 \maghundred\ in \textit{V}-band and duration of $ \sim77 $d) of a generic IIL SN. On the other hand \snhj\ also shows high plateau decline rate of $ 1.5 $ \maghundred\ in \textit{V}-band, similar to SNe IIL, but marginally lower than SNe IIL template light curves. Our high cadence photometric observations of \sne\ enables us to cover all characteristic phases up to radioactive tail of optical light curves. Broadband polarimetric observations reveal some polarization in \snhj\ with subtle enhancement as SN evolves towards plateau end, however the polarization angle remains constant throughout the evolution. This characteristic is consistent with the idea that the evolving SN with recombining hydrogen envelope is slowly revealing more asymmetric central region of explosion. Modelling of bolometric light curve yields a progenitor mass of $ \sim11\msun $ with a radius of $ \sim700\rsun $ for \snhj, while for \sng\ model estimated progenitor mass is $ \sim9\msun $ with a radius of $ \sim630\rsun $, both having a typical energy budget of $ \sim2\times10^{51} $ erg.

\end{abstract}


\section{Introduction} \label{sec:intro}

Core-Collapse SNe (CCSNe) originate from massive stars with $ M_{ZAMS}>8\msun $ \citep{2013RvMP...85..245B} which at the end of nuclear burning phase, having insufficient thermal energy start to collapse under self gravity. The massive core overcomes the electron degeneracy state following which process of neutronization takes place releasing large amount of energetic neutrinos which helps drive the explosion. CCSNe are sub classified into Hydrogen rich (Type II) and hydrogen deficient (Type I) SNe. The retention of significant amount of Hydrogen in SNe II at the time of explosion is visible as characteristic Balmer emissions in their spectra.

CCSNe are further divided in various sub types \citep[see][for a review of SN classifications]{1997ARA&A..35..309F}. Among these the most common sub-type is IIP which constitutes $ \sim60\% $ of all CCSNe \citep{2011MNRAS.412.1441L}. SNe IIP has extended hydrogen envelope which is left almost fully ionized at the time of shock breakout. As expanding envelope cools down, the ionized envelope starts to recombine with recombination front moving inward through the ejecta which is responsible for a sustained plateau of almost constant brightness for  $ \sim 80 - 100 $ days. At the end of recombination phase, SNe IIP experience rapid drop in luminosity settling onto a slow declining tail phase which is powered by gamma rays emitted from the radioactive decay of \cobalt\ to \iron\, which in turn depends upon the amount of short lived radioactive \nickel\ synthesized in explosion.

According to historical classification based on the light curve properties \citep{1979A&A....72..287B}, fast declining counterpart of type IIPs are SNe IIL, which are supposed to show almost a linear decline in the light curve until it reaches the tail phase, without any sort of flattening during early phases of light curve.
Over time it has been realized that such events are extremely rare that could fit in the criteria of SNe IIL.
Studies by \cite{2014ApJ...786...67A} and \cite{2015ApJ...799..208S} on a large sample of SNe II disapprove existence of any bimodality in light curve properties which can be grouped into distinct subclasses. Rather they found a continuum in light curve slopes and physical parameters. Moreover now it is also believed that a steep declining plateau-nebular transition like phase can be detected even in typical SNe IIL if they are observed long enough. The plateau slope of SNe II light curves are primarily governed  by the amount of hydrogen present in the ejecta. As in the case of SNe IIP where the hydrogen content is high enough with an extended envelope, the energy deposited  instantaneously from shock and also from  the initial decay of \nickel\ is released slowly over a longer period of time, which sustain the flat plateau phase. On the contrary, if the amount of intact hydrogen is relatively less, then energy will be released at a faster rate resulting in a fast declining light curve but with a brighter initial peak luminosity. Considering vast diversity of progenitor properties \citep{2009ARA&A..47...63S}, variable amount of hydrogen  mass intact in the ejecta would give rise to a continuum distribution of light curve slopes. \cite{2014MNRAS.445..554F} proposed a criteria to reestablish SNe IIL classification, according to which a SN with a decline of 0.5 mag in \textit{V}-band light curve during first 50 days would qualify as a type IIL. However this criteria is a bit too rigid for IIL classification as it will classify more than 50\% of all Type II SNe as Type IIL considering the sample of \cite{2014ApJ...786...67A}. Therefore, rather than using this criteria we use SNe IIL light curve template presented by \cite{2014MNRAS.445..554F} for the purpose of classification. It is also to be noted, that even by following scheme of template matching, many of the earlier studies which comprises of collective sample of SNe IIP may also include some reclassified SNe IIL as well.

Extensive theoretical and numerical attempts have been made to understand progenitor properties and relate them with other observable parameters of explosion \citep{1985SvAL...11..145L,1989ApJ...340..396A,2003ApJ...582..905H}. Stellar evolutionary models suggest that  these SNe can originate from red supergiant (RSG) stars of masses $ \sim9-25\msun $ \citep[e.g.][]{2003ApJ...591..288H}. However, progenitors directly identified from pre-SN archival \textit{HST} images for a decent number of IIP SNe suggest an upper mass limit of $ 16.5\pm1.5 $ \msun\ \citep{2009MNRAS.395.1409S}.
One of the many proposed explanation for the lack of high mass RSG progenitors is that stars above $ 17\msun $ may end up into IIL or IIn SNe \citep{2009ARA&A..47...63S}.

Extragalactic SNe are spatially unresolvable and hence its difficult to address structure and geometry of explosion. Polarimetric observations have proved to be a valuable probe in investigating asymmetry of SN explosion. In SNe the source of polarization is primarily the geometry of electron scattering atmosphere within the ejecta. Any asymmetry in this envelope would give rise to net polarization of observed light. CCSNe are often found to show significant degree of polarization in optical and IR wavelengths \citep{1996ApJ...462L..27W,2001ApJ...553..861L,2001PASP..113..920L,2002Msngr.109...47W,2006A&A...454..827P}. SNe with hydrogen envelope stripped (Type Ib/c) or partially stripped (Type IIb or IIn) exhibit higher level of polarization as compared to SNe with intact H envelope like SNe IIP (typically about 0.5 -- 0.8\% polarization). The thick H envelope obscures the observed asymmetry, whereas probing deeper towards the central part of the explosion, more polarization is observed implying higher degree of asymmetry in electron scattering atmosphere \citep{2005ASPC..342..330L}. However, in IIP SNe polarization enhancement is observed towards the end of plateau as SN starts to enter the nebular phase, where the hydrogen recombination is close to completion, and with decreasing opacity the inner asymmetric core is being revealed \citep[e.g. SN 2004dj;][]{2006Natur.440..505L}.
Generally  a moderate asphericity of $ \sim20\% $ would result in a linear polarization of $ \sim1\% $ \citep{2005ASPC..342..330L}.
Besides asphericity in electron scattering atmosphere, some amount of polarization may also originate from scattering by dust environment \citep{1996ApJ...462L..27W}, clumps in ejecta, asymmetric distribution of radioactive \nickel\ \citep{2006AstL...32..739C} or asymmetric ionization of outer envelope due to shock produced in CSM interaction.

The first broadband polarimetric observation was done for SN 1987A \citep{1988MNRAS.234..937B} demonstrating significant polarization and its time evolution. Late time \textit{HST} observations of spatially resolved ejecta revealed elongation in a direction consistent to the polarization angle inferred from polarimetry \citep{2002ApJ...579..671W}, which validated our understanding to use polarimetry as a probe for SN explosion geometry. SNe 1999em and 2004dj are extremely well studied and observed spectropolarimetrically, which provided detailed insight on the nature of these explosions \citep{2001ApJ...553..861L,2006Natur.440..505L}. While spectropolarimetry has advantage of wavelength resolved polarization information, on the other hand broadband polarimetry is useful to extract overall polarimetric information and infer any asymmetry without the requirement of bright SN or powerful observational resources.

In this work we present broadband photometric and  polarimetric observations of two fast declining type II \sne. A brief introduction of these events is given in Section~\ref{sec:sne.intro}. This paper is organized as follows. In Section.~\ref{sec:obs} we describe photometric and polarimetric observations and data reduction procedure. Adopted distance and estimated extinction values for both the SNe are discussed in Section~\ref{sec:ext}. In Section~\ref{sec:lc} we analyze photometric light curve properties of \sne\ and compare with other SNe II to establish its position in Type II diversity. In this section we also estimate \nickel\ mass from computed bolometric light curves.
The explosion parameters and progenitor properties are estimated by modelling the light curve, which are described in Section~\ref{sec:modelling}. Polarimetric analysis and its temporal evolution is discussed in Section~\ref{sec:polarimetry}. Finally in Section~\ref{sec:sum} we summarize the results obtained in this work.

\subsection{\sne} \label{sec:sne.intro}
\sne, both are moderately bright events which have been detected very young. \snhj\ was discovered on December 12.3 UTC, 2013 by \cite{2013CBET.3757....1A} in the galaxy \hosthj\ ($ \sim 30 $ Mpc). Unfortunately there is no reported non-detection prior to SN explosion, which can be used to observationally constrain the explosion epoch. The classification spectrum from ASIAGO on December 13.1 UTC shows broad and very shallow \ha\ and \hb\ P-Cygni profiles on top of featureless blue continuum. Comparison of this spectrum with well studied SNe templates, using GELATO \citep{2008A&A...488..383H} and SNID \citep{2007ApJ...666.1024B}, is found similar to type II SNe and matches best with SN 2012A spectrum at 2.6 day after explosion. Moreover, from visual comparison with other SNe II spectra \citep[e.g.,][]{2014MNRAS.438L.101V,2015MNRAS.450.2373B} at early epochs, the December 13 spectrum looks no older than four days post explosion. Therefore, we adopt an explosion epoch (0d) of December 10.5 UTC, 2013 ($ \rm JD= 2456637.0\pm1.5 $ days) for \snhj.

\sng\ was discovered on January 14.5 UTC, 2014 in \hostg\ ($ \sim 24 $ Mpc) and was reported in \textit{CBAT Transient Object Followup Reports}\footnote{http://www.cbat.eps.harvard.edu/unconf/tocp.html}. They also reported an earlier faint detection ($ \sim17.4 $ mag) in CCD image taken on January 13.6 UTC with limiting magnitude of 18.5. \cite{2014CBET.3787....2D} reported last confirmed non-detection of the SN on January 10.9 UTC, 2014 up to a limiting magnitude of 19.4. Thus we adopt mid time between the first faintest detection and last non-detection as the explosion epoch (0d) for \sng, which is January 12.2 UTC, 2014 ($ \rm JD=2456669.7\pm1.4 $ days). \cite{2014ATel.5767....1O} reported optical spectrum taken on January 15.1 UTC, showing \Hi\ line with narrow \ha\ emission  and unusually strong \Heii\ line, which was found similar to type IIn SN 2013cj. However, it was again reclassified as type IIL by \cite{2014ATel.5935....1E} from spectrum taken several weeks after peak. The basic parameters of these two SNe and their host galaxies are tabulated in Table.~\ref{tab:host}.

\input{./host.tex}

\section{Observation and Data reduction} \label{sec:obs}
\subsection{Photometry}\label{sec:photometry}
Broadband photometric observations of \sne\ have been carried out in \textit{UBVRI} bands. Rigorous follow up observations for both the SNe were initiated soon after their discovery.
The observations in the nebular phase continued until the objects went behind the Sun.
High cadence monitoring was done using ARIES 104cm Sampurnanand Telescope (ST) and 130cm Devasthal Fast Optical Telescope (DFOT) located at Nainital, India. Additional \textit{BVRI} data of \snhj\ was also collected from 50cm telescope at Osaka Kyoiku University, Japan. \textit{Swift} Ultraviolet optical telescope (UVOT) also observed \sng\ for nearly a month in all six UVOT bands.

\begin{figure}
\centering
\includegraphics[width=\linewidth]{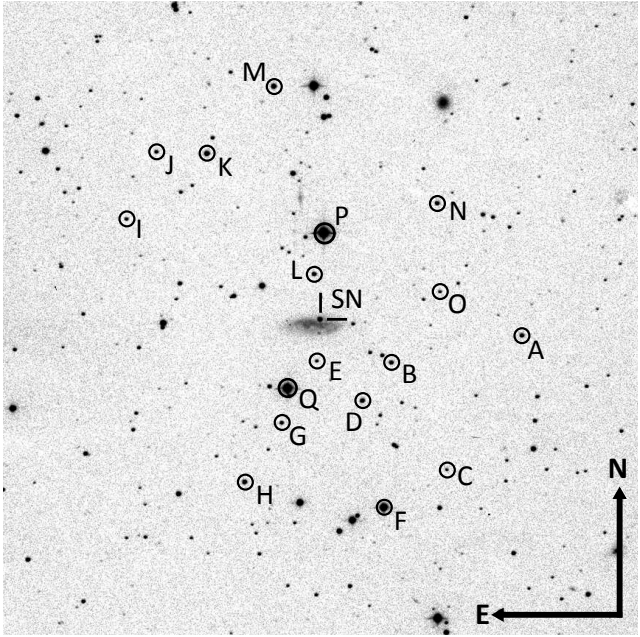}
\caption{\snhj\ in \hosthj. The $V$-band image taken from 104-cm ST covering a subsection of
         about 11\arcmin$\times$11\arcmin\ is shown. The secondary field standards are marked with circles.}
\label{fig:idsnhj}
\end{figure}

\begin{figure}
\centering
\includegraphics[width=\linewidth]{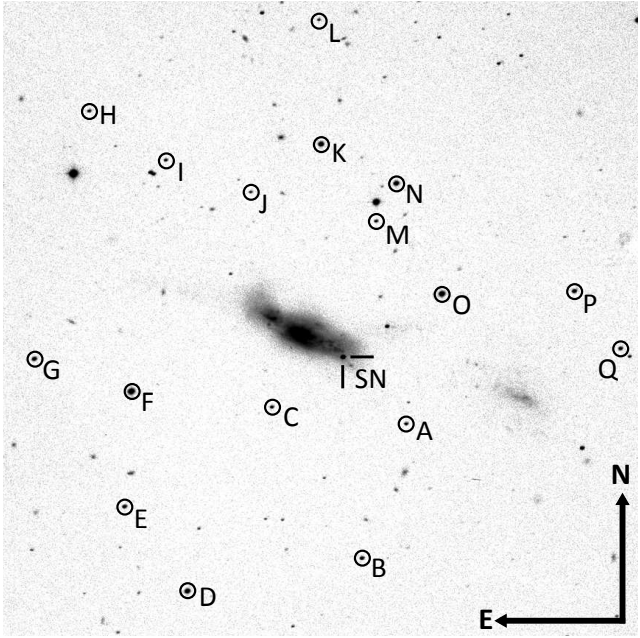}
\caption{\sng\ in \hostg. The $V$-band image taken from 104-cm ST covering a subsection of
         about 11\arcmin$\times$11\arcmin\ is shown. The secondary field standards are marked with circles.}
\label{fig:idsng}
\end{figure}

The routine data cleaning, reduction, extraction of light curve and field calibration to generate local standards follows the same procedure as outlined in  \cite{2013MNRAS.433.1871B}.
Fig.~\ref{fig:idsnhj} and Fig.~\ref{fig:idsng} shows the field of \sne\ respectively. As seen in the figures both the SNe are fairly embedded in their host galaxies and the estimated SN flux measurements will be contaminated by the background specially during late phases.
It is therefore necessary to correct for the host galaxy fluxes and in order to do so galaxy templates have been observed. Template images of the host galaxies in all five photometric bands were taken during 7 to 10 May 2015 when the SNe have fainted much below the detection limit (at an age of $ \sim515 $ and $ \sim480 $ days for \snhj\ and \sng\ respectively). In order to do the template correction,
we adopted aperture flux subtraction method, in which the SN flux and host galaxy background flux (from the template image) is extracted using the same aperture and at the same exact same location. The SN fluxes are then subtracted by the estimated background flux resulting into true SN flux free from galaxy contamination.
This entire process is achieved with a set of self written scripts which employs \daophot\ within it. In principle, the technique of template image subtraction should be adopted for best results \citep{1998ApJ...503..325A,2000A&AS..144..363A}. However, in this technique there are number of shortcoming with data obtained from small ground based telescopes. Stellar point spread function (PSF) may often be degraded in terms of minor elongation and skewness due to defect in passive telescope optics, which may further vary depending on the telescope's altitude-azimuth. In such cases, PSF matching becomes non trivial and may leave behind significant and arbitrary residual in flux, resulting into larger errors in photometry. In our case, such defects are prevailing in several images when the SNe were observed at high zenith angles.

To standardize both the SNe fields, four \cite{2009AJ....137.4186L} standard fields (PG 1323, PG 1633, SA 104 and SA 107) were observed on March 08 2014 from 104-cm ST under good photometric night and seeing ($ \rm FWHM\sim 2\arcsec $ in \textit{V} band) conditions. For determination of atmospheric extinction, multiple standard fields were observed with varying air masses. SNe fields were also observed during the same night, so that the derived transformation can be used to calibrate instrumental magnitudes of SNe. The obtained transformation equations with zero-point, color and atmospheric extinction coefficients are given as,
  \begin{eqnarray}
  u &=& U + (7.841\pm0.012) - (0.071\pm0.013) \cdot (U-B)  \nonumber\\ & & {} + (0.660\pm0.034) \cdot X \nonumber\\
  b &=& B + (5.319\pm0.010) - (0.072\pm0.010) \cdot (B-V)  \nonumber\\ & & {} + (0.402\pm0.020) \cdot X \nonumber\\
  v &=& V + (4.715\pm0.007) - (0.063\pm0.006) \cdot (B-V)  \nonumber\\ & & {} + (0.247\pm0.014) \cdot X \nonumber\\
  r &=& R + (4.429\pm0.007) - (0.057\pm0.010) \cdot (V-R)  \nonumber\\ & & {} + (0.259\pm0.024) \cdot X \nonumber\\
  i &=& I + (4.892\pm0.010) - (0.050\pm0.010) \cdot (V-I)  \nonumber\\ & & {} + (0.117\pm0.015) \cdot X
  \end{eqnarray}
where \textit{ubvri} are instrumental magnitudes corrected for time and aperture, \textit{UBVRI} are standard magnitudes and \textit{X} is the airmass. These transformation equations are applied to the SNe fields to calibrate 17 secondary standard stars in each of the fields of \sne. The secondary standards are marked  in Fig.~\ref{fig:idsnhj} and Fig.~\ref{fig:idsng}, and their calibrated magnitudes are listed in Table~\ref{tab:photstar}. The final photometry of \snhj\ and \sng\ are tabulated in Tables~\ref{tab:photsnhj} and \ref{tab:photsng} respectively.

\sng\ was also observed with UVOT \citep{2005SSRv..120...95R} on board \textit{Swift} spacecraft \citep{2004ApJ...611.1005G} in all six UVOT bands (\textit{uvw2, uvm2, uvw1, uvu, uvb, uvv}). The UVOT photometry was obtained from \textit{Swift} Optical/Ultraviolet Supernova Archive \citep[SOUSA; ][]{2014Ap&SS.354...89B}. The reduction procedure for UVOT photometry is based on \cite{2009AJ....137.4517B}, which includes subtraction of background galaxy counts, and adopting the revised zero-points and time dependent sensitivity from \cite{2011AIPC.1358..373B}. The UVOT photometry  for \sng\ is presented in Table~\ref{tab:photsng}.

\subsection{Polarimetry}
Broadband polarimetric observations in \textit{R}-band have been carried out for \sne\ using ARIES Imaging Polarimeter \citep[AIMPOL;][]{2004BASI...32..159R} mounted at 104cm ST. The unvignetted field of view of the CCD image is $ \sim8\arcmin $ in diameter.
Stellar FWHM is found to vary within 2 to 3 pixels, CCD read-noise and gain are 7.0\el and 11.98 \el/ADU respectively.

Bias correction and cosmic ray removal is done following usual procedure. The Wollaston prism splits light into ordinary and extraordinary rays and make it incident on the CCD. The corresponding stellar images are separated by $ \sim28 $ pixels along the north-south direction on the sky plane. Images are taken in four positions by rotating the half-wave plate through an angle $\alpha= 0\degr $, 22.5\degr, 45\degr\ and 67.5\degree, and the corresponding reduced stokes parameters are designated as $ q1 $, $ u1 $, $ q2 $ and $ u2 $ respectively and are determined as,
\begin{equation}
R(\alpha)=\frac{I_e(\alpha)/I_o(\alpha)-1}{I_e(\alpha)/I_o(\alpha)+1}=P\cos(2\theta-4\alpha).
\label{eq:stokes}
\end{equation}
Here, $ I_o $ and $ I_e $ are the intensities of ordinary and extraordinary stellar images determined from standard aperture photometry. $ P $ is the fraction of linear polarization and $ \theta $ is the angle of polarization plane. Correction factor for  nonuniform responsivity of (i) CCD pixels and (ii) system to orthogonal polarized beams, is also introduced as a ratio described by \cite{1998A&AS..128..369R},
\begin{equation}
\frac{F_o}{F_e}=\left[\frac{I_o(0\degr)}{I_e(45\degr)}\times
\frac{I_o(45\degr)}{I_e(0\degr)}\times
\frac{I_o(22.5\degr)}{I_e(67.5\degr)}\times
\frac{I_o(67.5\degr)}{I_e(22.5\degr)}   \right]^{0.25},
\end{equation}
which are multiplied with $ I_e/I_o $ ratios in Eq.~\ref{eq:stokes}. The obtained $ q1, u1, q2, u2 $ values are fitted with $ cosine $ function to obtain $ P $ and $ \theta $. In principle, a single set of $ q $ and $ u $ value is sufficient to determine polarization parameters, but all four stokes parameters are used as a redundancy check. The errors associated with stokes parameters are dominated primarily by photon statistics and is expressed as \citep{1998A&AS..128..369R},
\begin{equation}
\sigma_{R(\alpha)}=(\sqrt{N_e+N_o+N_{Be}+N_{Bo}})/(N_e+N_o),
\end{equation}
which are further propagated appropriately in deriving polarization parameters. Here $ N_o $, $ N_e $ are ordinary and extraordinary counts and $ N_{Bo} $, $ N_{Be} $ are ordinary and extraordinary background counts respectively for a position $ \alpha $.

Along with SNe, a number of polarized and unpolarized standards \citep{1992AJ....104.1563S}
have been also observed to estimate instrumental polarization and correct the zero-point polarization angle which mainly originate from imperfection in manual alignment of the instrument. For the polarized standards, the degree of polarization $ P $ measured from the instrument is found consistent within limit of errors and any offset found in the measured and standard values of polarization angle $ \theta $ is corrected in the measurements for SNe. Using unpolarized standards, the instrumental polarization is found to vary within 0.03 to 0.10\%, which is consistent to that found by other authors for AIMPOL \citep[see e.g.,][]{2004BASI...32..159R,2013A&A...556A..65E}. Since the direction of polarization is not confined to any particular plane and the value itself is very small (within error limits), we have not subtracted this from our observational measurements.

To get an estimate of the polarization towards the direction of \sne\ due to Galactic dust,we also observed a number of bright field stars (on April 22 2015) within an average radius of  $ \sim5\degr $ around corresponding targets. The $ V $-band polarization observations of these bright field stars are also presented in \cite{2000AJ....119..923H}.
We have observed 11 and 3 field stars in the direction of \sne\ respectively.

The observed polarization measurements for \sne\ are listed in Table~\ref{tab:pol.SNe} and those for field stars are tabulated in Table~\ref{tab:pol.star}.

\section{Distance and extinction} \label{sec:ext}
Distance and total line-of-sight extinction are the two parameters which are important to study the intrinsic properties of SNe. Thus it is essential to determine these parameters for each of the SNe, before we characterize and analyze them.

\subsubsection*{\snhj}
\hosthj\ hosts \snhj\ at the edge of the projected view of the diffused spiral galaxy. As there is no redshift independent distance available in the literature, we adopt a redshift distance of $ 28.2\pm2.0 $ Mpc from \texttt{Hyperleda}\footnote{http://leda.univ-lyon1.fr/} \citep{2014A&A...570A..13M} corrected for Virgo infall velocity and assuming $ H_0=70 $ km~s$ ^{-1} $Mpc$ ^{-1}$, $ \Omega_m=0.27 $, $ \Omega_\Lambda=0.73 $. However there are well accepted techniques which utilizes observations of type II SN itself to determine its distance. \epm\ \citep{1974ApJ...193...27K,2009ApJ...696.1176J,2014ApJ...782...98B} and \textsc{scm} \citep{2002ApJ...566L..63H,2010ApJ...715..833O} are two such methods which have been successfully implemented on several SNe II to estimate redshift independent distances, but requires extensive spectroscopic information as well which restricts us from implementing these methods on both of these SNe.

The most widely adopted method to estimate line-of-sight extinction is from correlation of narrow \Nai~D absorption line equivalent width with the reddening \ebv\ \citep[see ][]{1990A&A...237...79B,2003fthp.conf..200T,2012MNRAS.426.1465P}. Since our dataset is only limited to photometry, we can not adopt this method which requires SN spectrum of good resolution and high SNR.
In order to determine total extinction we need estimates for both Milky way and host galaxy extinction.
Using the all-sky Galactic extinction map by \cite{2011ApJ...737..103S}, we adopt a reddening value of \ebv$ _{\rm MW}=0.045\pm0.002 $ mag due to Milky Way.
To have an estimate of extinction due to host galaxy we implement the ``color-method" \citep{2010ApJ...715..833O}, which assumes that intrinsic $ (V-I) $ color towards the plateau end is constant. \cite{2010ApJ...715..833O} found an empirical relation between color and visual extinction as,
\begin{eqnarray}
A_V(V-I)&=&2.518[(V-I)-0.656] \nonumber\\
\sigma_{(A_V)}&=&2.518\sqrt{\sigma^2_{(V-I)}+0.053^2+0.059^2} \label{eq:oliv}
\end{eqnarray}
Putting a $ (V-I) $ color of $ 0.725\pm0.031 $ mag, which is a low-order cubic interpolated value at 75d and corrected for Galactic extinction, we derive the host galaxy visual extinction of $A_{V_{host}}=0.172\pm0.069 $ mag. Therefore, the adopted total line-of-sight visual extinction is $ A_V=0.312\pm0.214 $ mag, which translates to $ \ebv=0.100\pm0.069 $ mag assuming the ratio of total to selective extinction $ R_V=3.1 $.

\subsubsection*{\sng}
\sng\ is hosted in the galaxy \hostg. From previous studies, only a single redshift independent distance estimate is available using Tully-Fisher method \citep{1988ang..book.....T} for the galaxy, which is $ 24.4\pm9.0 $ Mpc. We adopt this distance for \sng\ throughout the paper.

The value of reddening due to Milky Way towards \hostg\ is found to be $\ebv_{\rm MW}= 0.010\pm0.000 $ mag from the all-sky Galactic dust-extinction survey \citep{2011ApJ...737..103S}. To estimate the host galaxy extinction for \sng, we implement the color-method \citep{2010ApJ...715..833O} as described in the case of \snhj. Following their prescription, using $ (V-I) = 0.956\pm0.038 $ mag (interpolated at 56 day after correcting for Galactic extinction) in Eq.~\ref{eq:oliv}, we obtain host visual extinction $ A_{V_{host}}=0.755\pm0.222 $ mag. Thus summing up, we adopt a total extinction of $ A_V=0.787\pm0.222 $ mag, corresponding to a reddening value $ \ebv=0.254\pm0.072 $ mag.

\section{Optical light curve} \label{sec:lc}

 \subsection{Apparent magnitude light curves} \label{sec:lc.app}

 \subsubsection*{\snhj}
\begin{figure}
 \centering
 \hspace*{-4.0mm}
 \includegraphics[width=1.05\linewidth]{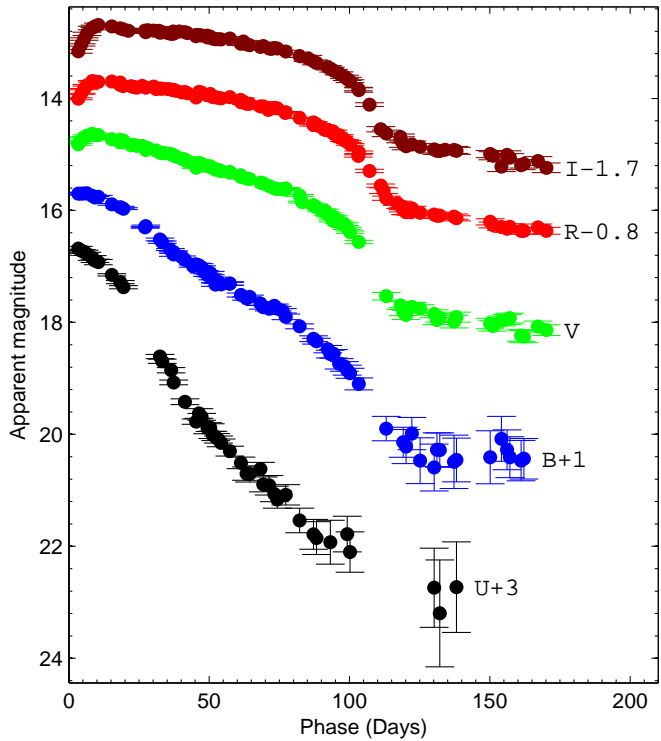}
 \caption{The photometric light curve of \snhj\ in Johnson-Cousins \ubvri\ bands. The light curves are shifted for clarity as indicated on the plot.}
 \label{fig:lc.app13hj}
 \end{figure}

 \textit{UBVRI} broadband photometric observations of \snhj\ are available at 117 phases from 3 to 170d. The data coverage is dense and almost continuous during this span. The resultant light curves are shown in Fig.~\ref{fig:lc.app13hj}.

 The early light curves show a sharp rise in \textit{RI} bands, which is also visible in \textit{BV} bands but relatively much flatter. After the initial peak the light curves decline slowly until the plateau ends. \snhj\ does not show very sharp transition in slope at the end of plateau as is seen in generic SNe IIP, e.g., SNe 1999em \citep{2002PASP..114...35L}, 1999gi \citep{2002AJ....124.2490L}, 2004et \citep{2006MNRAS.372.1315S}, 2013ab \citep{2015MNRAS.450.2373B}. Rather the light curve shows linearity only up to $ \sim80 $d, thereafter it shows a gradual increase in slope until it reaches the end of plateau at around $ \sim96-100 $d. At the end of plateau, the light curve declines steeply to settle onto a relatively slow declining radioactive tail at $ \sim118 $d, which continues till the end of our observations.

 The light curve parameters of \snhj\ are listed in Table~\ref{tab:lcpar} (Rows:1--3). The time of peak in early light curves (Tab.~\ref{tab:lcpar}, Row:1) are consistent with most typical and common fast-rising  SNe IIP like 2005cs \citep{2009MNRAS.394.2266P}, 2004et \citep{2006MNRAS.372.1315S} and 2013ab \citep{2015MNRAS.450.2373B}, which is significantly faster than slow-rising SNe, e.g. SNe 2006bp \citep{2007ApJ...666.1093Q}, 2009bw \citep{2012MNRAS.422.1122I} and 2012aw \citep{2013MNRAS.433.1871B}.
 After the initial peak the light curves continue to decline during the entire plateau phase, however in \textit{R} and \textit{I} bands a break in slope is visible at 23 and 24 days respectively. This break is possibly the mark of the emergence of recombination phase, as the shock heated ejecta is cooling down and recombination starts to power the light curve. Such a feature, but more prominent, was also seen in \textit{VRI} bands of type IIP SN 2012aw \citep{2013MNRAS.433.1871B}.

Since the initial peak \snhj\ continues to decline, but does not show a constant slope towards the plateau end. During 80 to 100d a gradual increase in light curve slope can be seen. The plateau slope for the linear part is listed in Tab.~\ref{tab:lcpar} (Row:2). This is steeper than the values reported for generic type IIP SNe 1999em \citep{2002PASP..114...35L}, 1999em \citep{2002AJ....124.2490L}, 2012aw \citep{2013MNRAS.433.1871B} and 2013ab \citep{2015MNRAS.450.2373B}, but flatter than the fast declining SNe like SN 2013by \citep{2015MNRAS.448.2608V}, SN 2013ej \citep{2015ApJ...806..160B} and 2014G (this paper).
For example, decline rates for SN 2012aw are 5.60, 1.74, 0.55 mag (100 d)$ ^{-1} $ in $ UBV $-bands and for SN 2013ab decline rates are 7.60, 2.72, 0.92, 0.59 and 0.30 mag (100 d)$ ^{-1} $ in \textit{UBVRI} bands, whereas SN 2013ej decline rates are 6.60, 3.57, 1.74, 1.07 and 0.74 \maghundred.
Decline rates for radioactive tail phase in \textit{VRI} bands are listed in Tab.~\ref{tab:lcpar} (Row:3), where \textit{U} and \textit{B} band slopes were not computed due to large errors in tail data points.

 \begin{table*}
 \centering
 \caption{Light curve parameters of \snhj\ and \sng.}
\renewcommand{\arraystretch}{1.6}
 \label{tab:lcpar}
  \begin{tabular}{ccl|ccccccccccc}
  \hline \hline
 &Row No. & Parameter & \textit{U} & \textit{B} & \textit{V} & \textit{R} & \textit{I} & \textit{uvw2} & \textit{uvm2}  & \textit{uvw1} & \textit{uvu} & \textit{uvb} & \textit{uvv} \\ \hline
\multirow{3}[0]{*}{\begin{sideways}\snhj\end{sideways}}
 & 1 & Rise time of early peak (days)           & --  & 6 & 8 & 11 & 12 & --  & --  & --  & --  & --  & --  \\
 & 2 & Plateau slope [\maghundred] & 5.81 & 3.38 & 1.50 & 0.89 & 0.77 & --  & --  & --  & --  & --  & --  \\
 & 3 & Nebular slope [\maghundred] & --  & --  & 0.72 & 0.81 & 0.76 & --  & --  & --  & --  & --  & --  \\ \hline
\multirow{3}[0]{*}{\begin{sideways}\sng\end{sideways}}
 & 4 & Rise time of early peak (days)           & 6 & 10 & 13 & 15 & 17 & --  & --  & 5 & 6 & 10 & 11  \\
 & 5 & Plateau slope [\maghundred] & 6.66 & 4.09 & 2.55 & 1.97 & 1.83 & 18.63 & 20.40 & 15.33 & --  & --  & --  \\
 & 6 & Nebular slope [\maghundred] & --  & 1.40 & 1.57 & 1.52 & 1.78 & --  & --  & --  & --  & --  & --  \\ \hline
 \end{tabular}
\end{table*}

\subsubsection*{\sng}
\begin{figure*}
\centering
\hspace*{-4.0mm}
\includegraphics[width=0.6\linewidth]{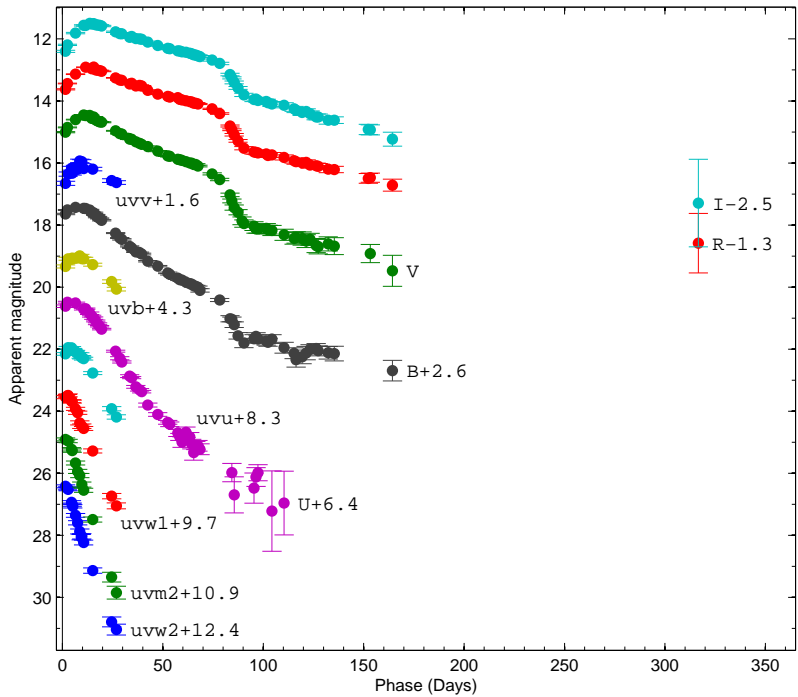}
\caption{The photometric light curve of \sng\ in Johnson-Cousins \ubvri\ and \swift~UVOT bands. The light curves are shifted for clarity as indicated on the plot.}
\label{fig:lc.app14g}
\end{figure*}

Photometric measurements in \textit{UBVRI} bands for \sng\ were made at 63 phases during 4 to 319d, where continuous and high-cadence observation were only during 4 to 166d, with a late nebular data point at 319d recovered only after galaxy background subtraction. Additionally 15 phases of UVOT observations during 2 to 27d  is also available in all six UVOT bands (\textit{uvw1, uvw2, uvm2, uvu, uvb} and \textit{uvv}). The light curves are shown in Fig.~\ref{fig:lc.app14g}.

The early part of light curves show a sharp rise in most of the optical bands (\textit{BVRI} and \textit{uvb, uvv}). The rising is detectable in UV bands as well  except for \textit{uvm2} and \textit{uvw2} bands. After the initial peak all the light curves decline linearly until plateau end ($ \sim77 $d), like a generic SNe IIL (SN 1980K; \citealt{1982AA...116...35B} and SN 2013by; \citealt{2015MNRAS.448.2608V}).  From $ \sim77 $d, the light curve decline steeply to enter relatively slow declining radioactive tail phase starting from $ \sim90 $d.

The light curve parameters for \sng\ are tabulated in Table~\ref{tab:lcpar} (Rows:4--6). Early rising of \sng\ light curve (Tab.~\ref{tab:lcpar}, Rows:4) is consistent with slow rising SNe, e.g, the \textit{V}-band maximum is delayed to 16d in SN  2006bp \citep{2007ApJ...666.1093Q}, 13d in SN 2009bw \citep{2012MNRAS.422.1122I} and 15d in SN 2012aw \citep{2013MNRAS.433.1871B}. This is significantly slower than most typical type II SNe, like SNe 2005cs, 2004et, 2013ab or the other SN which we present here - \snhj.

After reaching maxima, \sng\ light curves in all bands shows linear decline throughout the plateau and their decline rates are listed in Tab.~\ref{tab:lcpar} (Rows:5). These values are similar to that of type IIL SNe like SN 1980K and  SN 2013by, but much steeper than all type IIP SNe (decline rates as mentioned above for SNe 2012aw and 2013ab). For example, SN 2013by \citep{2015MNRAS.448.2608V} shows a recline rate of 3.62, 2.01, 1.42 and 1.26 \maghundred\ in \textit{BVri} bands and 19.53, 20.54 and 15.26 \maghundred\ in \textit{uvw2, uvm2, uvw1} bands. The decline rates for radioactive tail light curves are also listed Tab.~\ref{tab:lcpar} (Rows:6). The slope of radioactive tail light curve is also found to be steeper than normal IIP SNe (e.g. SN 2013ab tail slopes are 0.36, 0.97, 0.66, 1.16 \maghundred\ in \textit{BVri} or in SN 2012aw tail slopes are 0.88, 0.88, 0.81, 0.95 \maghundred\ in \textit{BVRI}) but very similar to that found in SN 2013ej \citep[tail slopes are 1.22, 1.53, 1.42, 1.55 \maghundred\ in \textit{BVRI} bands]{2015ApJ...806..160B}.

\subsection{Absolute magnitude light curves}

\begin{figure*}
\centering
\includegraphics[width=16cm]{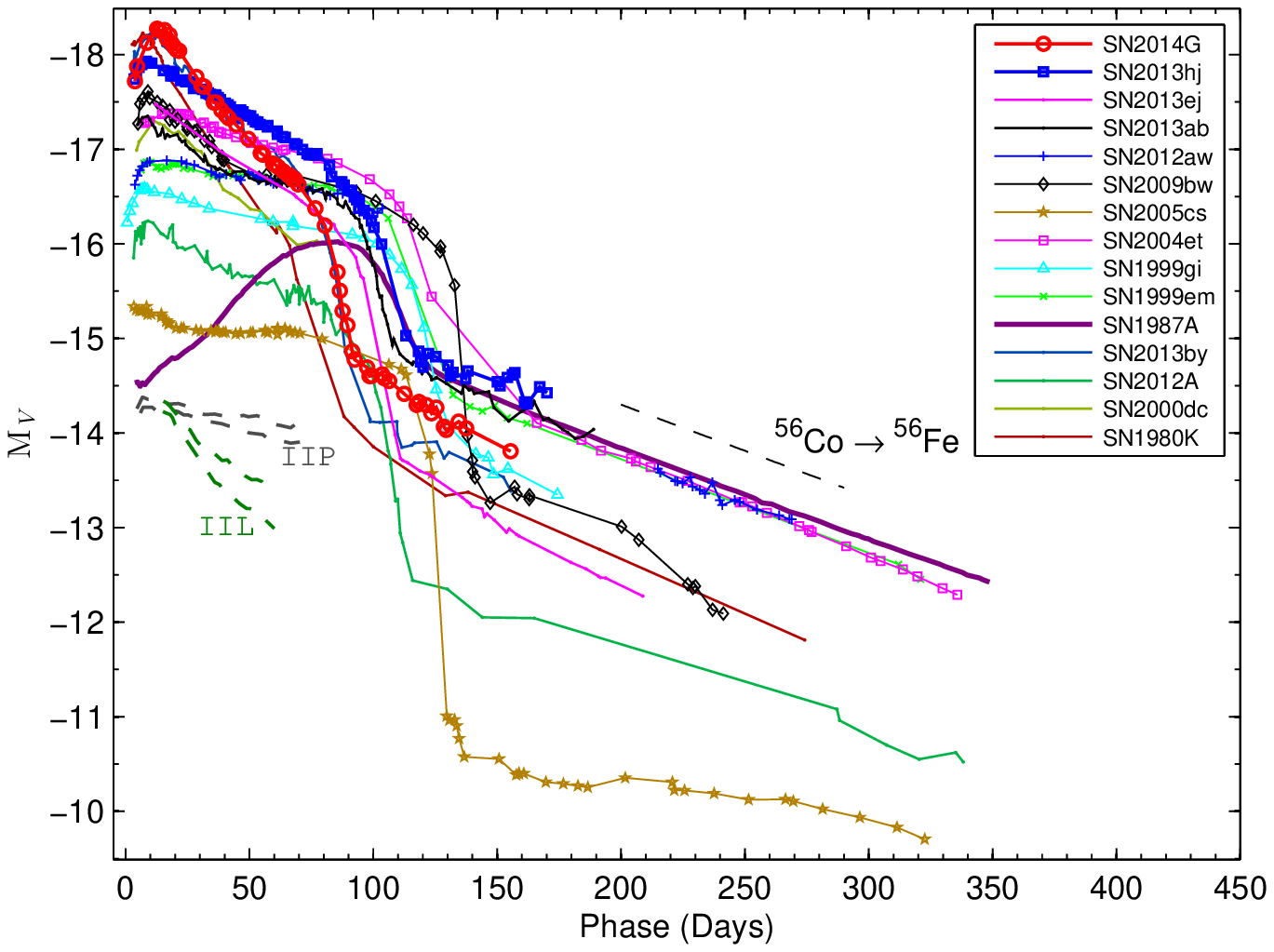}
\caption{
Absolute \textit{V} band light curves of \sne\ are compared with other type II SNe.
The exponential decline of the tail light curve following the radioactive decay of \cobalt$ \rightarrow $\iron\ is shown with a dashed line. On the bottom left side, two pair of dotted lines in gray and green colors represent the slope range for type IIP and IIL SNe templates as presented by \citet{2014MNRAS.445..554F}.
The adopted JD, distance, \ebv\ and reference for V-band magnitude are listed in Table~\ref{tab:adoptedparms}.
  }
\label{fig:lc.abs}
\end{figure*}

 \begin{table*}
 \caption{Adopted parameters and references for $ V $-band magnitudes for the SNe II sample used in this paper.}
   \label{tab:adoptedparms}
 \begin{tabular}{lcccl}
 \hline\hline
  SNe       & JD      & Distance & \ebv &  Reference for data \\
            & 2400000+& (Mpc)    & (mag)&            \\ \hline
  SN 1980K  & 44540.5 & ~5.5 & 0.30 & \citet{1982AA...116...35B}; NED database$ ^a $         \\
  SN 1987A  & 46849.8 & 0.05 & 0.16 & \citet{1990AJ.....99.1146H}                       \\
  SN 1999em & 51475.6 & 11.7 & 0.10 & \citet{2002PASP..114...35L,2003MNRAS.338..939E}   \\
  SN 1999gi & 51522.3 & 13.0 & 0.21 & \citet{2002AJ....124.2490L}                       \\
  SN 2000dc & 51762.4 & 49.0 & 0.07 & \citet{2014MNRAS.445..554F}; NED database$ ^a $         \\
  SN 2004et & 53270.5 & ~5.4 & 0.41 & \citet{2006MNRAS.372.1315S}                       \\
  SN 2005cs & 53549.0 & ~7.8 & 0.11 & \citet{2009MNRAS.394.2266P}                       \\
  SN 2009bw & 54916.5 & 20.2 & 0.31 & \citet{2012MNRAS.422.1122I}                       \\
  SN 2012A  & 55933.5 & ~9.8 & 0.04 & \citet{2013MNRAS.434.1636T}                       \\
  SN 2012aw & 56002.6 & ~9.9 & 0.07 & \citet{2013MNRAS.433.1871B}                       \\
  SN 2013ab & 56340.0 & 24.0 & 0.04 & \citet{2015MNRAS.450.2373B}                       \\
  SN 2013by & 56404.0 & 14.8 & 0.19 & \citet{2015MNRAS.448.2608V}                       \\
  SN 2013ej & 56497.3 & ~9.6 & 0.06 & \citet{2015ApJ...806..160B}                       \\ \hline
 \end{tabular}
 \begin{flushleft}
 $ ^a $ \url{http://ned.ipac.caltech.edu}
 \end{flushleft}
 \end{table*}

\input{./slopendrop}

Absolute \textit{V}-band ($ M_V $) light curves of \sne\ are shown in Fig.~\ref{fig:lc.abs} and are compared with a moderately sized sample of SNe II. In order to determine light curve characteristics of these two SNe and mark their position in the type II diversity, we selected the comparison sample with diverse range of properties, comprising of archetypal type IIP SNe 1999em, 1999gi, 2012aw to prototypical type IIL SN 1980K. The plateau slopes for all SNe II sample shown in Fig.~\ref{fig:lc.abs} are listed in Table.~\ref{tab:slopendrop}. This table and figure is primarily an adaptation from \cite{2015ApJ...806..160B}.

The mid plateau absolute magnitude for \snhj\ is highest amongst the compared sample of events. However, it may be noted that the uncertainty in adopted reddening and distance values may alter this inference to some extent. The plateau decline rate of \snhj\ is higher than most generic IIP SNe (e.g., SN 1999em, SN 2005cs, SN 2012aw) but lower than IIL SNe 1980K or 2000dc. According to \cite{2014MNRAS.445..554F}, type II SN with plateau decline of at least 0.5 mag in first 50 days would be classified as type IIL, which would definitely classify \snhj\ as  type IIL. However, this criteria proposed by \cite{2014MNRAS.445..554F} is a bit too rigid, as it would reclassify all SNe up to slopes of SN 2012A (see Table~\ref{tab:slopendrop}) as type IIL. \cite{2014MNRAS.445..554F} also presented template light curve range for SNe IIP and IIL, which is shown in Fig.~\ref{fig:lc.abs}.
The plateau slope for \snhj\ is marginally smaller than the minimum slope for the SNe IIL templates but it is much larger than the steepest slope for the SNe IIP templates.
Thus, if \snhj\ has to be classified either between type IIP or IIL, it would fit best in the class of type IIL rather than type IIP. The closest comparison to \snhj\ in terms of plateau slope, shape and duration is SN 2013ej. However, plateau decline of SN 2013ej is a bit more steep than \snhj\ making it compatible to SNe IIL light curve templates.

\begin{figure}
 \centering
 \includegraphics[width=1.02\linewidth]{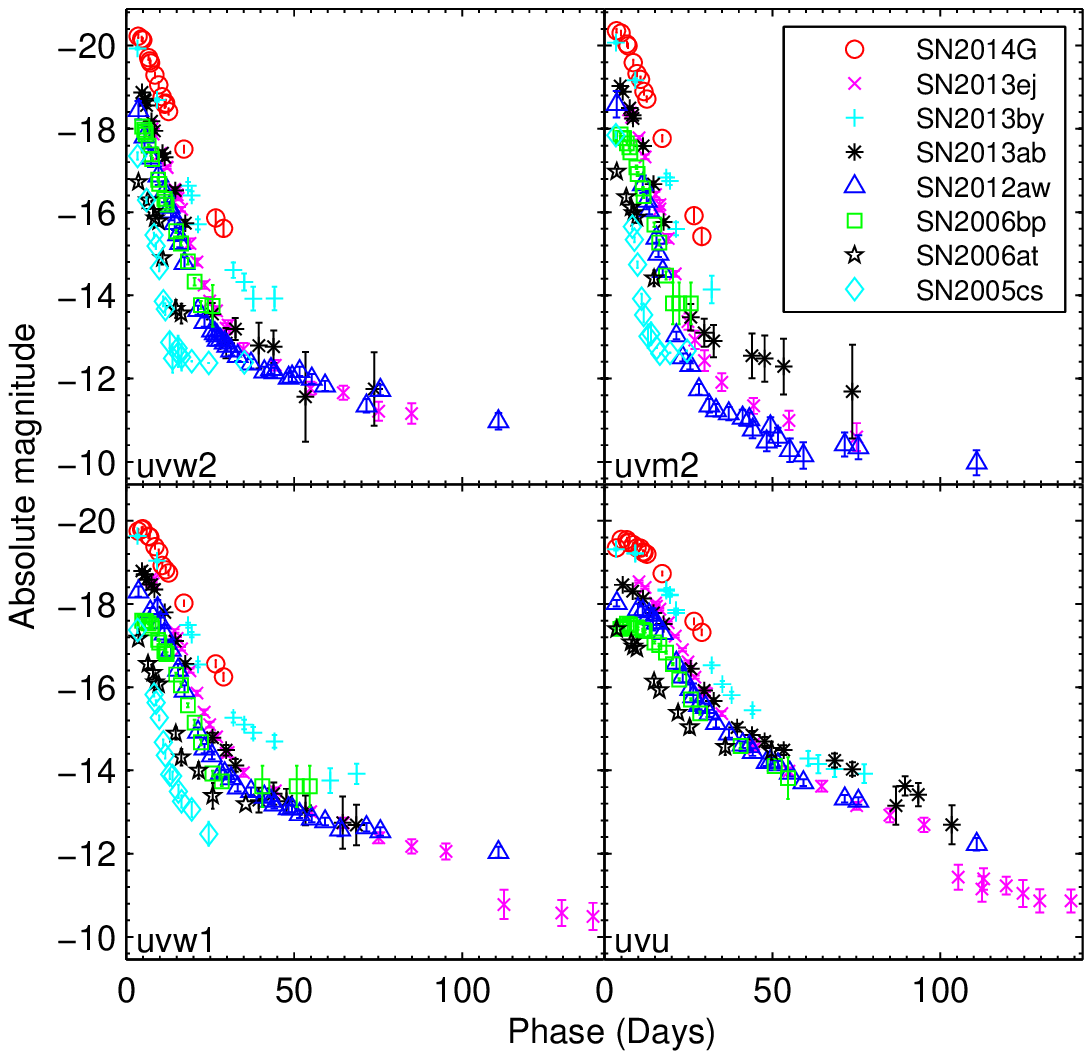}
 \caption{\swift~UVOT UV absolute light curves of \sng\ is compared with other well observed SNe II from UVOT. For the compared SNe, references for UVOT data, extinction and distance are: SN 2005cs -- \citet{2009AJ....137.4517B,2009MNRAS.394.2266P}, SN 2006at -- \citet{2009AJ....137.4517B}; Distance 65 Mpc; $ \ebv=0.031 $ mag \citep[only Galactic reddening][]{2011ApJ...737..103S}, SN 2006bp -- \citet{2008ApJ...675..644D}, SN 2012aw -- \citet{2013ApJ...764L..13B,2013MNRAS.433.1871B}, SN 2013by -- \citet{2015MNRAS.448.2608V}, SN 2013ej -- \citep{2015ApJ...806..160B}.}
 \label{fig:uv.abs}
 \end{figure}

On the other hand, owing to the fast decline of \sng, its mid-plateau magnitude is lower than \snhj, but the peak luminosity is highest amongst the compared sample. The decline rate is higher than all SNe II in the sample, except for prototypical type IIL SN 1980K and SN 2000dc. The plateau slope of \sng\ is marginally over the SNe IIL template, thus clearly qualifying as a type IIL SN. SN 2013by is the closest  comparison to \sng\ in terms of light curve shape and magnitudes except for the fact that \sng\ plateau is steeper and the nebular tail is brighter. The decline rate of tail light curve is 1.57 \maghundred, which is higher than that expected for light curve powered from radioactive decay of \cobalt\ to \iron. However, such steepening of tail light curve was also seen in other type IIL SNe as well. For example, SN 2013ej exhibits a steeper decline rate of 1.53 \maghundred\ \citep{2015ApJ...806..160B}, similarly other IIL SNe 1980K and 2013by also show steeper declines like \sng\ (see Fig.~\ref{fig:lc.abs}).
In Fig.~\ref{fig:uv.abs} we show the \textit{Swift} UVOT absolute magnitude light curves for \sng\ in UV bands and are compared with other SNe II which are well observed by UVOT. UV light curve evolution of \sng\ is found similar to other SNe. The peak brightness closely matches to that of SN 2013by but is higher than most other SNe in the sample.

Based on the light curve parameters, \sng\ is a perfect example of a generic type IIL SN whereas \snhj\ belongs to an intermediate class with plateau slope close but lower than typical SNe IIL.
With the adopted reddening and distance values, both the SNe lie on the luminous end of the comparison sample, which may eventually fall in the range of normal luminosity if reddening or distance values are lowered. \cite{2014ApJ...786...67A} found an anti-correlation between the plateau slope and duration, which is consistent with the trend as we see for \sne\ as well as other SNe in the sample.

 \subsection{Bolometric light curve} \label{sec:lc.bol}

 \begin{figure*}
 \centering
 \includegraphics[width=0.8\linewidth]{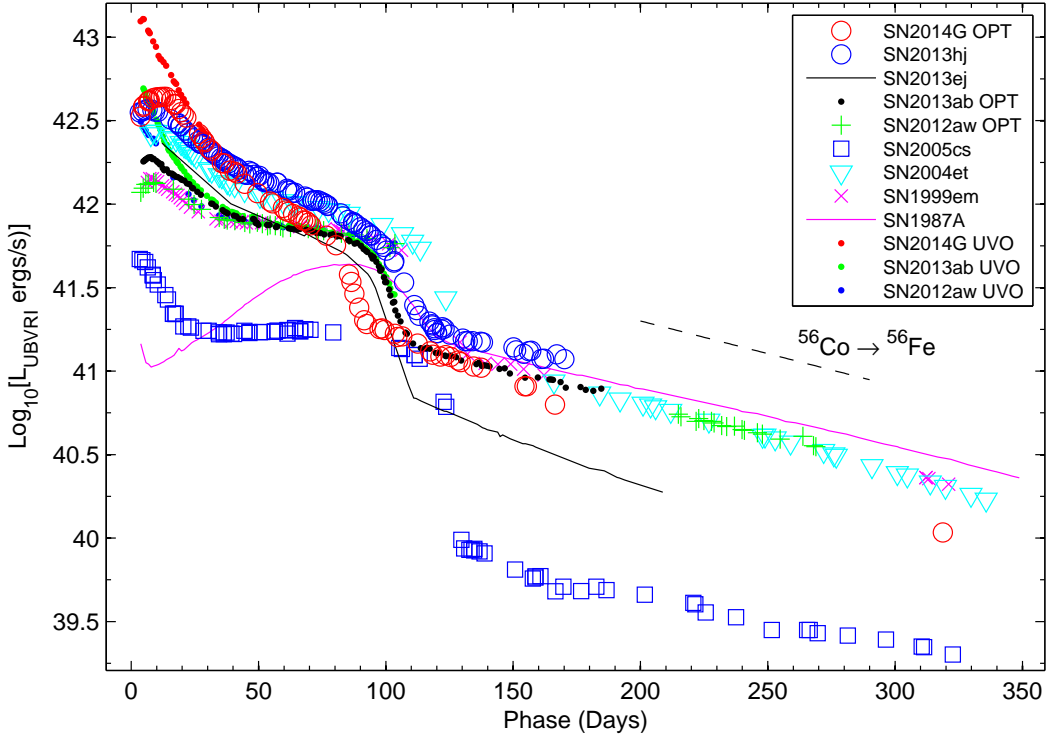}
 \caption{The \ubvri\ pseudo-bolometric light curves of {\sne} are compared with those of other well studied SNe. Light curves with added \textit{Swift} UVOT UV contributions are also shown for SN 2013ej, SN 2013ab and SN 2012aw (labeled as UVO).
          The adopted distances, reddening and time of explosion values are listed in Table~\ref{tab:adoptedparms}. The
          exponential decline of the radioactive \cobalt\ decay law is shown with a dashed line.}
 \label{fig:lc.bol}
 \end{figure*}

Pseudo-bolometric light curves for \sne\ are computed using photometric fluxes corrected for adopted reddening and distance values.
We follow the same method as outlined in \cite{2013MNRAS.433.1871B} to compute bolometric luminosities, which involves semi-deconvolution of broadband filter response from photometric fluxes. Thermal emission from SNe mostly peaks in optical and near-optical wavelengths depending on the phase of SN evolution. At early phases ($ <30 $d) when the SN  is hot, UV wavelengths dominate bolometric luminosity. Likewise, at late phases ($ >100 $d) when the SN is cool enough, bolometric light is dominated by infrared fluxes. We compute pseudo-bolometric luminosity for both the SNe in the optical domain, which include \textit{UBVRI} fluxes. For \sng\ we also computed bolometric luminosity by taking UV fluxes into account, which include fluxes from \textit{uvw2} to I bands. The UV+Optical luminosity yields a significantly higher value at early times. In Fig.~\ref{fig:lc.bol} we compare bolometric light curves of \sne\ with other well-studied SNe. The bolometric light curves of the comparison sample were also computed using the same method and within the same wavelength range.

Bolometric light of \snhj\ suffers almost a linear decline during 6 to 100d with a luminosity drop of $ \sim0.85 $ dex. This decline rate is similar to SN 2013ej, but higher than most generic IIP SNe in the sample. An additional drop of 0.47 dex is seen at the end of plateau while light curve settles onto the radioactive nebular tail phase since 118d. The slope of tail light curve is similar to those of SNe 1987A, 1999em, 2004et and 2013ab which is consistent to that expected for \cobalt\ to \iron\ radioactive decay rate.

\sng\ shows a decline rate higher than all compared SNe including \snhj. \sng\ shows a linear decline throughout the plateau phase and experiences a drop of 0.89 dex in bolometric luminosity during 13 to 80d.
A drop of 0.49 dex in luminosity is seen during the transition from plateau to nebular phase.
\sng\ light curve settles onto relatively slow declining radioactive tail phase after 92d. The slope during this phase is found to be somewhat steeper (0.39 dex $ (100)^{-1} $d) than other generic IIP SNe (e.g. SNe 1987A, 1999em, 2004et, 2013ab), and also that expected for a light curve powered by radioactive decay of \cobalt\ to \iron. On carefully examination of the absolute \textit{V}-band and bolometric light curve (see Figs.~\ref{fig:lc.abs} and \ref{fig:lc.bol}), such steepening of tail light curve is also seen in other type IIL SNe 1980K, 2013by and 2013ej. Steepening of tail light curve would indicate inefficient trapping of gamma rays in the ejecta. In type IIL SNe, the hydrogen envelope is relatively more depleted than IIP counterpart, resulting into lower gamma ray optical depth causing leakage of gamma photons. Such a scenario may explain steeper tail light curve in \sng, and this may also be true for all other fast declining SNe IIL as well. However, due to fast decaying brightness, there are only a handful of SNe IIL
with observations extending up to the tail phase and \sng\ is the newest among such candidates.

 \subsection {Mass of nickel} \label{sec:lc.nick}

In CCSNe, radioactive \nickel\ is produced by explosive nucleosynthesis at the time of explosion. The nebular phase light curve is mainly powered by the radioactive decay chain of \nickel\ to \cobalt\ and \cobalt\ to \iron, with $ e $-folding life time of 8.8 and 111.26 d respectively. Thus, the tail luminosity would be proportional to the amount of radioactive nickel synthesized at the time of explosion.

SN 1987A is one of the most well studied SN with a fair degree of accuracy in the estimation of \nickel\ mass  \citep[$ 0.075\pm0.005\msun $;][]{1980ApJ...237..541A}.
By comparing the tail luminosity of a SN with that of SN 1987A, we can estimate the mass of \nickel\ for that SN as well. Although in principle true bolometric luminosities (including UV, optical and IR) are to be used to compute \nickel\ mass, but we do not have IR fluxes for either of \snhj\ or \sng. For the uniformity in comparison we also use \textit{UBVRI} bolometric luminosity for SN 1987A, computed using the same method and wavelength range.
We compute the ratio of \snhj\ to SN 1987A luminosity at 150d to be $ 1.152\pm0.168 $. This corresponds to a \nickel\ mass of $M_{Ni}({\rm 2013hj})= 0.086\pm0.013 $ \msun. Similarly, the ratio of \sng\ to SN 1987A luminosity at 150d is $ 0.735\pm0.552 $, which corresponds to $M_{Ni}({\rm 2014G})= 0.055\pm0.041 $ \msun.

The gamma photons emitted from the radioactive decay of \cobalt\ to \iron\ are assumed to thermalize the ejecta powering the tail light curve, which in turn depends on the initial mass of synthesized \nickel. \cite{2003ApJ...582..905H} related these parameters as,
 \begin{eqnarray*}
  M_{\rm Ni} = 7.866\times10^{-44} \times L_{t} \exp\left[ \frac{(t_{t}-t_{0})/(1+z)-6.1}{111.26}\right]\msun,
 \end{eqnarray*}
 where $t_{0}$ is the explosion time, 6.1d is the half-life of \nickel\ and 111.26d is the e-folding time of the \cobalt\ decay. Using the bolometric correction factor for tail phase as given by \cite{2003ApJ...582..905H}, we compute a tail luminosity of $ 2.71\pm0.20\times 10^{41} $ erg at 150d for \snhj,  which correspond to \nickel\ mass of $M_{Ni}({\rm 2013hj})= 0.077\pm0.010 $ \msun. Likewise, tail luminosity of \sng\ at 140d is $ 1.78\pm1.01\times 10^{41} $ erg  corresponding to $M_{Ni}({\rm 2014G})= 0.045\pm0.035 $ \msun.

The amount of \nickel\ masses estimated from both the methods are consistent with each other within errors. Thus we adopt a mean value of both the results, which is $M_{Ni}({\rm 2013hj})= 0.080\pm0.008 $ \msun\ for \snhj,  and $M_{Ni}({\rm 2014G})= 0.050\pm0.027 $ \msun\ for \sng.
Although the quoted errors in \nickel\ mass include uncertainty from adopted reddening and distance values, but does not take care of selection bias involved in these parameters. Estimated \nickel\ masses may be lowered if lower values of reddening or distance is assumed.

\section{Light curve modelling}\label{sec:modelling}
We model the bolometric light curves of \sne\ following the semi-analytical approach originally developed by \cite{1980ApJ...237..541A} and further refined  in \cite{1989ApJ...340..396A}. Such simplistic models \citep[e.g.][]{1980ApJ...237..541A,1982ApJ...253..785A,1989ApJ...340..396A,1993ApJ...414..712P,2003MNRAS.338..711Z,2012ApJ...746..121C} are useful to get a preliminary yet reliable estimate of explosion parameters without running more accurate but resource intensive hydrodynamical models \citep[e.g.][]{1977ApJS...33..515F,2007A&A...461..233U,2011ApJ...729...61B,2011ApJ...741...41P}. The model has been implemented by \cite{2014A&A...571A..77N} and \cite{2015ApJ...806..160B} for several type II SNe to estimate explosion parameters and are found to be fairly consistent with results from hydrodynamical models.

Description of model formulation and algorithm is presented in \cite{2015ApJ...806..160B} and references therein. The temporal component of temperature evolution in a co-moving frame of expansion is given \citep{1989ApJ...340..396A,2014A&A...571A..77N} as,
 \begin{equation}
 \frac{d\phi(t)}{dz}= \frac{R(t)}{R_0 x_i^3}\left[p_1\zeta(t)-p_2\phi(t)x_i-2 x_i^2 \phi(t) \frac{R_0}{R(t)}\frac{dx_i}{dz}\right] ,
 \end{equation}
which is numerically solved with appropriate treatment of dimensionless recombination front $ x_i $. After finding the solution for $ \phi(t) $ and $ x_i $, the model luminosity is computed as \citep{1989ApJ...340..396A,2014A&A...571A..77N,2015ApJ...806..160B},
\begin{equation}
 L(t)=x_i\frac{\phi(t)E_{th}(0)}{\tau_d}\left(1-e^{-A_g/t^2}\right)+4\pi r_i^2 Q\rho(x_i,t)R(t)\frac{dx_i}{dt} .
\end{equation}
The parameters in these equations have standard meaning as referred in aforementioned papers.
$ A_g $ is the effectiveness of gamma ray trapping {\citep[see e.g., ][]{1997ApJ...491..375C,2012ApJ...746..121C}} which is an important parameter to model a radioactive tail steeper than that expected for light curve powered by decay chain of \nickel$ \rightarrow $\cobalt$ \rightarrow $\iron.

 \begin{figure}
 \centering
 \hspace{-0.5cm}
 \includegraphics[width=1.05\linewidth]{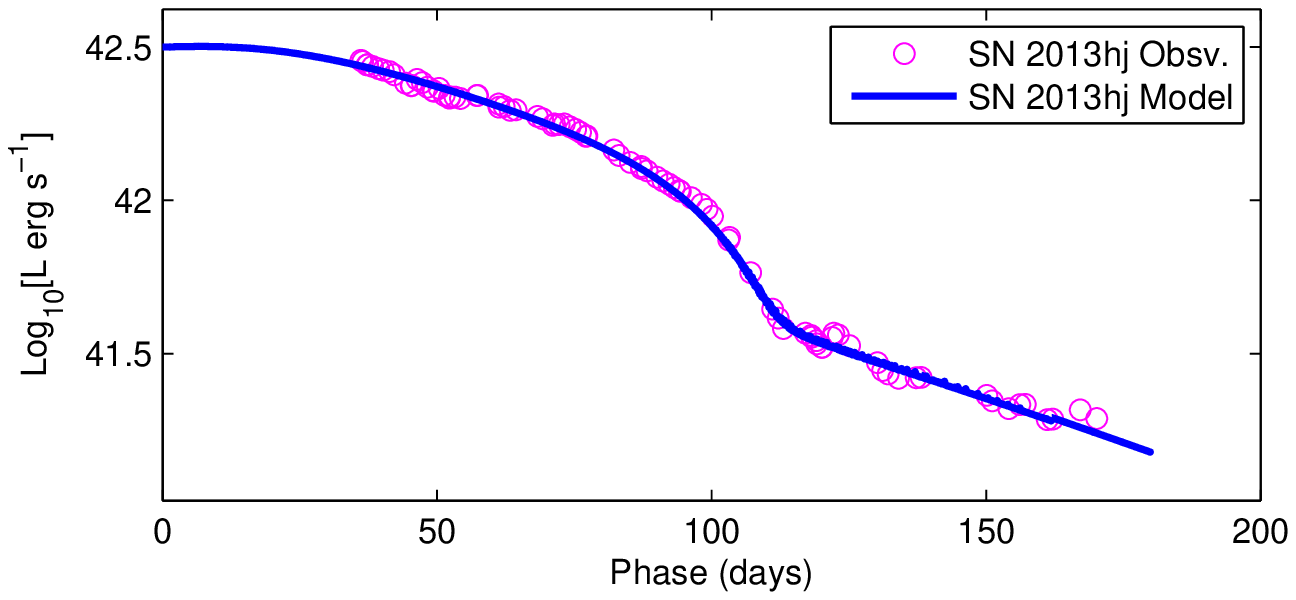}\\
  \hspace{-0.5cm}
 \includegraphics[width=1.05\linewidth]{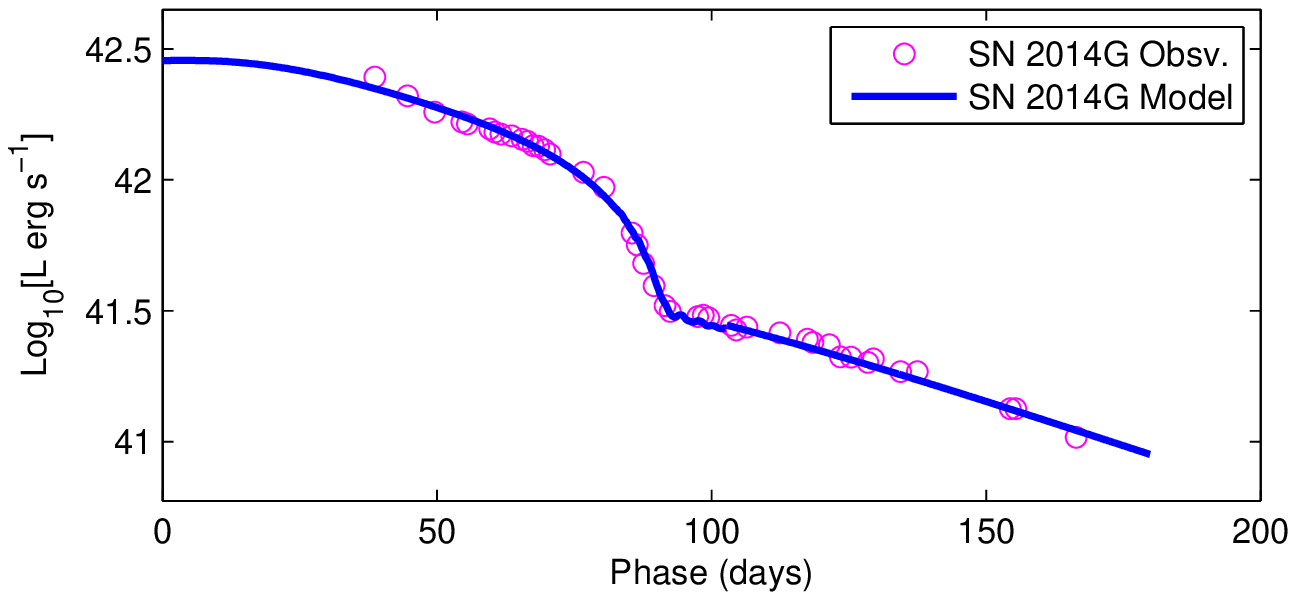}
 \caption{The observed bolometric light curves of \sne\ are shown with the best fit model.}
 \label{fig:model}
 \end{figure}

To implement the model, we require true bolometric luminosities for the SNe. Since our data is limited only to UV and optical, we add IR flux contribution assuming \sne\ have similar optical to IR flux ratio as observed for SN 1999em  at similar phases. Fig.~\ref{fig:model} shows the computed bolometric light curves of \sne\ fitted with our model. Since these models are inefficient to reproduce post-breakout peaks of early light curves, we restricted the data only from later part of photospheric phase.

From the best fit model of \snhj\ we estimate an ejecta mass of 9.6 \msun, progenitor radius of 700 \rsun\ and total explosion energy $ \sim2.1 $ foe. The mass of radioactive \nickel\ incorporated in the model is 0.075 \msun, which is consistent to the value estimated in Section~\ref{sec:lc.nick}. Assuming the mass of compact remnant to be 1.5--2.0 \msun, the total progenitor mass adds up to $ \sim11 $ \msun.

Similarly for \sng, model estimated ejecta mass is 7.0 \msun, progenitor radius is 630 \rsun\ and total explosion energy is $ \sim2.1 $ foe. \nickel\ mass determined from model is 0.052 \msun\ which is almost equal to the value estimated in Section~\ref{sec:lc.nick}. Adding up a mass of 1.5 -- 2.0 \msun\ for the compact remnant, the total progenitor mass is $ \sim9 $ \msun.
To fit the steeper tail light curve of \sng, as already discussed in Sections~\ref{sec:lc.app} and \ref{sec:lc.bol}, we increased the gamma ray leakage in the model by lowering the effectiveness of gamma ray trapping parameter $ A_g $ to $ 4\times10^4 \rm day^2$. This parameter is related to gamma-ray optical depth as $ \tau_g\sim A_g/t^2 $.

\section{Broadband polarimetry} \label{sec:polarimetry}
\begin{figure}
 \centering
 \hspace{-0.5cm}
 \includegraphics[width=8.8cm]{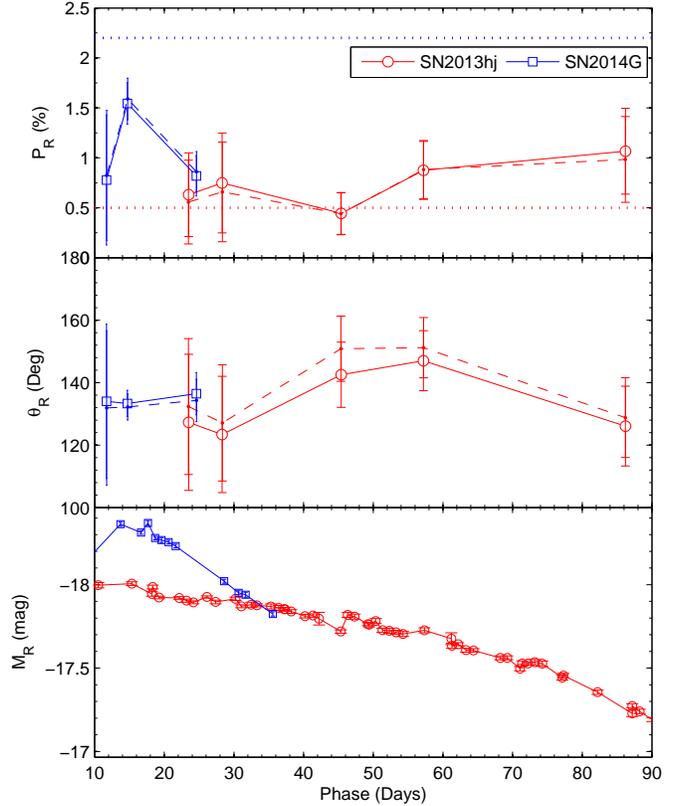}%
 \caption{The degree of polarization $ P $, polarization angle $ \theta $ and absolute $ R $-band magnitude $ M_R $ is plotted in three panels for \sne. The dashed lines in polarization plots represent the observed polarization parameters, whereas the solid connected plots represent polarization parameters after subtracting for $ \rm ISP_{MW} $. The horizontal dotted lines in top panel represents maximum amount of host galaxy polarization \citep[$ P_{\rm max}<9\times\ebv $;][]{1975ApJ...196..261S} component that could be present in SN polarization values (0.50 and 2.20 \% for \sne\ respectively; see Sec.~\ref{sec:host_isp}) which may be linearly subtracted if the polarization angles of host and SN are co-aligned.  \textit{R}-band light curves are truncated to only show phases where polarimetric data exists.}
 \label{fig:pol.lc}
 \end{figure}

Broadband polarimetric observations in $ R $-band has been carried out for \snhj\ at 5 phases during 23 to 86d, whereas for \sng\ only 3 phases of observation during 12 to 25d are available. The observed temporal evolution of $ P $ and $ \theta $ (dashed lines indicate measurement that have not been corrected for any ISP) for both the SNe, along with $ R $-band absolute light curve is plotted in Fig.~\ref{fig:pol.lc} (data is tabulated in Table.~\ref{tab:pol.SNe}).

\subsection{Estimation of interstellar polarization}
The observed polarization in supernova light primarily has three components, intrinsic polarization due to SN itself, interstellar polarization due to Milk Way ($ \rm ISP_{MW} $) and interstellar polarization due to host galaxy ($\rm ISP_{HG} $).
Thus, in order to interpret the intrinsic polarization of SN, we need to subtract ISP components from the observed polarization. However, there is no completely reliable method to derive ISP and thus posses an issue in SN polarimetric studies. Number of methods have been proposed to estimate ISP \citep{1993ApJ...414L..21T,1997PASP..109..489T,2001ApJ...550.1030W}, but all involve some assumptions which are not always valid \citep{2008ARA&A..46..433W}. Having this limitation, best possible attempt has been made to have a reasonable estimate of ISPs which are discussed in following sections.

\subsubsection{Milky way}
To get an estimate of $ \rm ISP_{MW} $, we observed 11 bright stars towards the direction of \snhj, within a maximum radius of $ 8\degr $ around the SN.
The \textit{V}-band polarization measurements of these stars are available in \cite{2000AJ....119..923H} catalog.
Similarly, field stars have also been observed towards the direction of \sng, but due to scarcity of bright stars only 3 have been selected (within a maximum radius of $ 13\degr $) also having polarization measurements in \cite{2000AJ....119..923H}. All these selected stars have very low \textit{V}-band polarization values listed \citep{2000AJ....119..923H}. The measured \textit{R}-band polarization (see Table \ref{tab:pol.star}) for these stars was also found to be low and of similar order as that of the \textit{V}-band. To compute the median ISP from observed stars, averaging is done on stokes components and which is then converted back to polarization parameters. The median of stokes components are computed as,
\begin{eqnarray}
\left<Q\right>&=&\left<~P_i\cos (2\theta_i) ~  \right>_{i=1..n} \nonumber \\
\left<U\right>&=&\left<~P_i\sin(2\theta_i) ~  \right>_{i=1..n}
\end{eqnarray}
which are then converted back to polarization parameters,
\begin{eqnarray}
P_{\rm MW} &=&  \sqrt{\left<Q\right>^2 + \left<U\right>^2} \nonumber \\
\theta_{\rm MW} &=& \frac{1}{2}\tan^{-1}\left(\frac{\left<U\right>}{\left<Q\right>}\right).
\end{eqnarray}

Thus, we obtain median $\rm ISP_{MW} $ as $ P_{\rm MW}=0.13\pm0.06 \% $ and $ \theta_{\rm MW}=12.4\pm10.3\degr $ for \snhj, while for \sng, $ P_{\rm MW}=0.08\pm0.22 \% $ and $ \theta_{\rm MW}=106.6\pm68.5\degr $. These $\rm ISP_{MW} $ values are then vectorially subtracted from the observed SN polarization which are also listed in Table~\ref{tab:pol.SNe} and plotted in (connected solid lines) Fig.~\ref{fig:pol.lc}.

According to \cite{1975ApJ...196..261S} the interstellar polarization efficiency due to Galactic dust is highly correlated to line-of-sight reddening, which may vary from 3 to 9 times the reddening value. As a reasonable approximation we adopt the mean polarization efficiency as $ P_{\rm mean}=5\times \ebv $. Using this relation and Galactic reddening values adopted (refer Section~\ref{sec:ext}) towards the direction each SN, we get $ P_{\rm mean} $ value of $ \sim0.23\% $ and $ \sim0.05\% $ for \sne\ respectively. These are roughly consistent with the $\rm ISP_{\rm MW} $ values we computed from observations of field stars, which implies that Galactic ISM follows the mean polarization efficiency.

\subsubsection{Host galaxy} \label{sec:host_isp}
For an accurate estimation of $ \rm ISP_{HG} $, one needs to know dust grain properties and magnetic field orientation along the line of sight within the host galaxy. In context to few SNe studies, it has been found that size of dust grains in respective host galaxy were dissimilar to Galactic dust grains (see e.g. SN 1986G; \citealt{1987MNRAS.227P...1H} and  SN 2001el; \citealt{2003ApJ...591.1110W}). Due to very diverse nature of dust grains in various galaxies, it is non trivial to adopt mean polarization efficiency without the knowledge of grain properties.
However, in order to get an idea of $ \rm ISP_{HG} $ and for the sake of understanding its possible implication in SN polarization measurements, we assume dust grain properties of galaxies \hosthj\ (for \snhj) and \hostg\ (for \sng) are similar to Galactic dust.
Now, we may estimate the maximum degree of polarization due to host galaxies following the \cite{1975ApJ...196..261S} relation, $ P_{\rm max}<9\times \ebv $.
In Section~\ref{sec:ext} we estimated host galaxy extinctions for both the SNe using color-method, which translates to \ebv$ _{\rm host} $ values of 0.055 and 0.244 mag for \sne\ respectively, assuming $ R_V=3.1 $.
Therefore, the corresponding $ P_{\rm max} $ (for host) values are $ <0.50\% $ and $ <2.20\% $ for \sne\ respectively.

To vectorially remove ISP components from SN polarization one need to know the direction information as well. It is generally believed and also found to be true that galactic magnetic fields run along the spiral arms \citep{1990IAUS..140..245S,1991MNRAS.249P..16S,2009IAUS..259..455H}, which is roughly perpendicular to the line joining galactic center and position of spiral arm \citep[see e.g., ][]{2014MNRAS.442....2K,2001ApJ...553..861L,2007ApJ...671.1944M}. However, in case of both \sne\ the host galaxies are almost irregular, not in face-on projection and SN location is not well resolved, this makes it difficult to make any reasonable estimate of the magnetic field orientation and effective line of sight polarization angle at SN locations.

Despite emphasizing all the critical aspects of $ \rm ISP_{HG} $, we do not subtract it from SN polarization measurements and make any speculative inference based on that. This was done mainly because of three uncertainties, (a) the reddening estimation of host galaxy is quite uncertain and may suffer systematic bias due to the method itself (see Section \ref{sec:ext}), (b) unknown grain properties in host galaxy may significantly vary the maximum polarization efficiency we are adopting and (c) without knowing the polarization angle of host galaxy vectorial subtraction will not be possible as a subtraction of ISP component may increase or decrease polarization values depending on $ \theta $. Thus, we caution the reader while interpreting the quoted polarization values as well as Figures~\ref{fig:pol.lc} and \ref{fig:pol.lc.all} in the paper.

\subsection{Intrinsic polarization and its evolution}
\begin{figure}
  \centering
  \hspace{-0.5cm}
  \includegraphics[width=8.8cm]{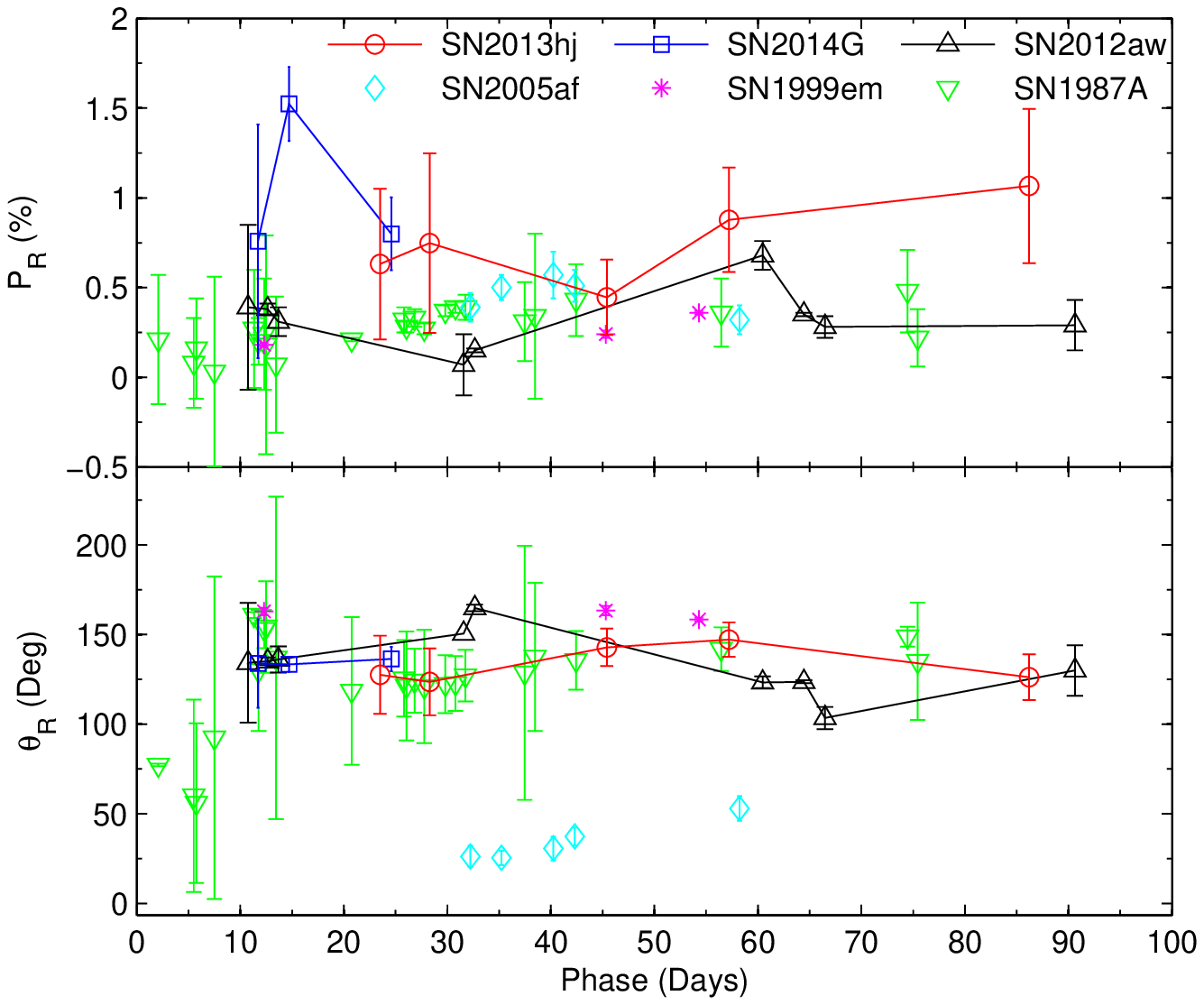}
  \caption{Broadband polarimetric evolution of \sne\ (only $ \rm ISP_{MW} $ subtracted values) are compared with other type IIP SNe. The references for data and explosion epochs are: SN 1987A -- \citet{1988MNRAS.234..937B}, SN 1999em -- \citet{2001ApJ...553..861L,2002PASP..114...35L}, SN 2005af -- \citet{2006A&A...454..827P,2005IAUC.8484....2F} and SN 2012aw -- \citet{2014MNRAS.442....2K}. }
  \label{fig:pol.lc.all}
  \end{figure}

The evolution of degree of polarization and the polarization angle of \sne\ are represented in Fig.~\ref{fig:pol.lc}.
The evolution of polarimetric parameters of both the events are also compared with other well sampled SNe IIP as shown in Fig. \ref{fig:pol.lc.all}.
Here SN 1999em observations are \textit{V}-band polarimetry, however temporal evolution in \textit{R}-band is expected to be similar. In both the figures the polarization values for \sne\ are subtracted for $\rm ISP_{MW} $ only, thus these measurements have components for intrinsic SN polarization as well as $\rm ISP_{HG} $. In Fig.~\ref{fig:pol.lc} horizontal dotted lines are shown to represent  the maximum degree of host polarization component which may be removed if orientation of host and measured SN polarization are assumed to be co-aligned. For any other $ \theta $ of $\rm ISP_{HG} $, the component of $ P $ due to $\rm ISP_{HG} $ in observed light will be lower than this level and so the intrinsic SN polarization will increase.

It is to be noted in Fig.~\ref{fig:pol.lc} that there is a small, yet significant amount of intrinsic polarization is left in \snhj\ even after considering maximum possible subtraction of $\rm ISP_{HG} $. About $ 0.3-0.5\% $ of polarization can be detected above maximum $\rm ISP_{HG} $ during 57 and 86d, which is well above the error bars of individual data points.  The $ P $ and $ \theta $ evolution does not show a significant relative variation among individual data points and is all within the limit of errors. However, a subtle increasing trend is noticed in overall $ P $ evolution, but $ \theta $ remains almost constant throughout.  This may imply that there is some asymmetry in electron scattering SN envelope, which is increasing according to envelope size and decrease in optical depth is revealing the asymmetric central part of explosion, but its geometrical orientation remains constant over time. It may be noted that the possible increase in polarization with time is almost linearly anti-correlated with the decline of SN brightness. Such monotonic increase in polarization without change in polarization angle has been also observed in SN 1999em (see Fig.~\ref{fig:pol.lc.all}) which was reported by \cite{2001ApJ...553..861L,2005ASPC..342..330L}.

\sng\ has only three measurements within first 25d. If maximum amount of $\rm ISP_{HG} $ is considered then entire detection is possibly submerged within the $\rm ISP_{HG} $ (see Fig.~\ref{fig:pol.lc}). The polarization angle for $\rm ISP_{HG} $ would play an important role in this case, only after knowing which we may conclude about the intrinsic polarization or asymmetry in SN envelope or CSM. Any favorable misalignment of  $\rm ISP_{HG} $ angle may lead to significant residual SN polarization. A sudden and unusual increase in polarization in noticed  at 15d which is coincident with the peak $ R $-band light, although no change in polarization angle is observed. However, considering the large errors and low density of observations, we restrict ourselves from making any conclusive inference based on a single measurement which may be resulting from a spurious detection.

\section {Summary} \label{sec:sum}

In this paper we present high-cadence photometric observation and a few epochs of broadband polarimetric observation of two fast declining type II \sne.

Due to lack of reliable handle on distance and reddening estimation, we adopted distance from single Tully fisher measurement or redshift distance available in literature, whereas host galactic reddening is estimate using `color-method' which itself is not very reliable. Details of these methods and adapted values are discussed in Section~\ref{sec:ext}. Implications of these uncertainties on the derived parameters and properties must be considered while making any interpretation. Although the derived parameters include the uncertainty from reddening and distance values, but it does not take care of the possible selection bias which might have  been introduced while adopting those values.

The plateau slope of \snhj\ is significantly steeper ($ 1.5 $ \maghundred\ in \textit{V}) than typical SNe IIP, but is similar to generic SNe IIL templates \citep{2014MNRAS.445..554F}.
\snhj\ is found to lie at an intermediate position in the broad continuum of light curve properties of SNe II diversity \citep[see][]{2014ApJ...786...67A}. If at all \snhj\ has to be classified in IIP/L taxonomy, then it will qualify better as a type IIL. \snhj\ does not show a sharp change in slope at plateau end ($ \sim98 $d) as we see in typical SNe IIP/L, rather a gradual increase of slope is observed. On the other hand, \sng\ is a perfect IIL candidate showing high initial peak brightness and steep plateau slope (2.55 \maghundred\ in \textit{V}) which lies at the higher end of SNe IIL light curve template. A sharp end of plateau phase is visible at $ \sim 77 $d for \sng.

Owing to fast decline of brightness, most IIL SNe do not have post-plateau observations. \sng\ is among the very few SNe IIL which has extensive observations covering all the light curve phases until nebular tail. The plateau-nebular transitional drop in luminosity is also prominent for the SN which is although expected but rarely detected for SNe IIL \citep{2014ApJ...786...67A}. Radioactive tail of \sng\ is found to be steeper than that expected for a light curve powered by radioactive decay chain of \nickel$ \rightarrow $\cobalt$ \rightarrow $\iron. Such steepening is explained by inefficient trapping of gamma rays in the ejecta. The tail light curves of type IIL SNe 1980K, 2013by and 2013ej also shows a significant steepening.
This possibly implies
that all type IIL SNe may exhibit such steepening of tail light curve, as SNe IIL has thinner hydrogen envelope which is responsible for gamma ray leakage and incomplete thermalization of photons. More type IIL SNe are needed to be observed rigorously until nebular phase to confirm whether steepening of tail is generic feature to all SNe IIL. The mass of synthesized radioactive \nickel\ estimated from tail bolometric light curves are  $ 0.08\pm0.01 $\msun\ and $ 0.05\pm0.03 $\msun\ for \sne\ respectively.

We performed semi-analytical modelling of bolometric light curves of \sne\ to estimate explosion and progenitor properties. For \snhj\, we estimate a progenitor mass of $ \sim11\msun $ with a radius of $ \sim700\rsun $, whereas for \sng\ progenitor mass is estimated to $ \sim9\msun $ with radius $ \sim630\msun $. The total explosion energy (kinetic + thermal) for both the SNe is approximately $ \sim2 $ foe. The steeper tail light curve of \sng\ is fitted by reducing the effectiveness of gamma ray trapping parameter $ A_g $ in the model.

Broadband polarimetric observations of \sne\ are also presented in this paper.
Even after considering maximum possible contamination in polarization due to host galaxy, \snhj\ is found to reveal some intrinsic polarization in its light (specially during mid- to late-plateau phases), which may become significant on adopting lower host reddening or by changing host polarization angle. The temporal evolution of polarization shows subtle but monotonic increment as SN evolves towards plateau end. It is generally expected for stripped or partially stripped SNe, namely Type Ib/c or IIb, to exhibit higher degree of polarization as compared to SNe with hydrogen envelope intact as in IIPs. As both of our studied SNe are fast declining Type II events, some fraction of H envelope may have already been shredded during pre-SN evolution in the form of stellar winds, thus the explosion will have relatively less amount of hydrogen retained which will quickly start to reveal the central part having higher degree of asymmetry \citep{2005ASPC..342..330L}. Evidence of similar polarization enhancement in SNe IIP during plateau to early nebular phase has been also reported by \cite{2001ApJ...553..861L,2006Natur.440..505L} for SNe 1999em and 2004dj.
On the other hand, polarization detected for \sng\ in our observations is of very low significance, and it may be within the level of interstellar polarization due to the host galaxy.

\section*{Acknowledgments}
We are thankful to the observers and observing staffs of ARIES and observatory of Osaka Kyoiku University for kind cooperation in observations of these objects. SB also acknowledges Dr. Eswaraiah C. for useful discussion on AIMPOL observations and data.  We gratefully acknowledge the services of the
NASA ADS, NED and SOUSA which are used to access data and references in this paper. We also thank the referee for the useful comments and suggestions on the manuscript.

\bibliography{ms}

\appendix
\section{Tables of photometry}
\input{./photstar}

\clearpage
\input{./photsnhj.tex}
\input{./photsng.tex}

\clearpage
\section{Tables of polarimetry}
\input{./polsn}

\input{./polstar}

\label{lastpage}

\end{document}

%% file: host.tex
  \begin{table}
  \caption{Parameters of SNe and their host galaxies.}
  \label{tab:host}
\setlength{\tabcolsep}{2.5pt}
  \begin{tabular}{llc} \hline \hline
     \noalign{\smallskip}
      Parameters& Value& Ref.\\
     \noalign{\smallskip} \hline
     \multicolumn{3}{c}{\bf\snhj}\\
     \noalign{\smallskip}
     \multicolumn{3}{l}{\textit{\underline \hosthj}:}\\
     Alternate name& PGC 025938 & 2\\
     Type& SBcd& 2\\
     RA (J2000)& $\alpha = 09^{\rm h} 12^{\rm m} 06\fs7$& 2\\
     DEC (J2000)& $\delta = -15\degr 25\arcmin 51\farcs9$& 2\\
     Abs. Magnitude& $M_{B}=-18.26$ mag& 2\\
     \\
     Distance& $D=28.2\pm2.0$ Mpc& 2 \\
     Distance modulus& $\mu = 32.25\pm0.15$ mag&\\
     \\
     Heliocentric Velocity& $cz_{\rm helio}=2060\pm7 \kms$&2\\
     \\
     \multicolumn{3}{l}{\textit{\underline \snhj}:}\\
     RA (J2000)& $\alpha = 09^{\rm h} 12^{\rm m} 06\fs3$& 3\\
     DEC (J2000)& $\delta = -15\degr 25\arcmin 46\farcs0$& \\
     \\
     Galactocentric Location& 6\farcs5 W, 5\farcs9 N&  3\\
     \\
     Time of explosion & $t_{\rm 0} =$ 10.5 December 2013 (UTC)& 1\\
                    & (JD $ 2456637.0\pm1.5 $ day)& \\
     Total reddening & \ebv\,= $0.100\pm0.069$ mag& 1\\
     \hline

	 \multicolumn{3}{c}{\bf\sng}\\
      \noalign{\smallskip}
      \multicolumn{3}{l}{\textit{\underline \hostg}:}\\
      Type& Sab& 2\\
      RA (J2000)& $\alpha = 10^{\rm h} 54^{\rm m} 39\fs4$& 2\\
      DEC (J2000)& $\delta = 54\degr 18\arcmin 18\farcs8$& 2\\
      Abs. Magnitude& $M_{B}=-20.07$ mag& 2\\
      \\
      Distance& $D=24.4\pm9.0$ Mpc& 2 \\
      Distance modulus& $\mu = 31.94\pm0.80$ mag&\\
      \\
      Heliocentric Velocity& $cz_{\rm helio}=1372\pm8 \kms$&4\\
      \\
      \multicolumn{3}{l}{\textit{\underline \sng}:}\\
      RA (J2000)& $\alpha = 10^{\rm h} 54^{\rm m} 34\fs1$& 5\\
      DEC (J2000)& $\delta = 54\degr 17\arcmin 56\farcs9$& \\
      \\
      Galactocentric Location& 44\arcsec W, 20\arcsec N&  5\\
      \\
      Time of explosion & $t_{\rm 0} =$ 12.2 January 2014 (UTC)& 1\\
                     & (JD $ 2456669.7\pm1.4 $ day)& \\
      Total reddening & \ebv\,= $0.254\pm0.072$ mag& 1\\

     \noalign{\smallskip}
     \hline
  \end{tabular}
  \newline (1) This paper;
        (2) HyperLEDA - http://leda.univ-lyon1.fr \citep{2014A&A...570A..13M};
  		(3) \citet{2013CBET.3757....1A}; (4) \citet{1988ang..book.....T};
  		(5) \citet{2014CBET.3787....2D}
  \end{table}

%% file: slopendrop.tex
\begin{table}
  \centering
  \setlength{\tabcolsep}{3pt}
  \caption{Parameters estimated from {\it V} band light cruve}
  \label{tab:slopendrop}
  \begin{tabular}{lccc } \hline \hline

       SN Name     &Plateau slope$^{a}$ &Transition drop$^{b}$&Transition time$^{c}$\\
                   &mag (100 d)$ ^{-1} $&mag                  &days           \\ \hline
		SN1980K    & 3.63 $\pm$ 0.04    &  2.0$\pm$0.2       & 37  $\pm$  5  \\
		SN2000dc   &~2.56 $\pm$ 0.06$^i$&  --                 & --            \\
		SN2014G    & 2.55 $\pm$ 0.02    &  1.6$\pm$0.2       & 17  $\pm$  2  \\
		SN2013by   & 2.01 $\pm$ 0.02    &  2.2$\pm$0.2       & 19  $\pm$  5  \\
		SN2013ej   & 1.74 $\pm$ 0.08    &  2.4$\pm$0.1       & 21  $\pm$  3  \\
		SN2013hj   & 1.50 $\pm$ 0.02    &  1.5$\pm$0.2       & 22  $\pm$  5  \\

		SN2012A    & 1.12 $\pm$ 0.03    &  2.5$\pm$0.1       & 23  $\pm$  4  \\
		SN2009bw   & 0.93 $\pm$ 0.04    &  2.4$\pm$0.2       & 14  $\pm$  3  \\
		SN2004et   & 0.73 $\pm$ 0.02    &  2.1$\pm$0.2       & 27  $\pm$  6  \\
		SN2013ab   & 0.54 $\pm$ 0.02    &  1.7$\pm$0.1       & 25  $\pm$  2  \\
		SN2012aw   & 0.51 $\pm$ 0.02    &   --                & --            \\
		SN1999gi   & 0.47 $\pm$ 0.02    &  2.0$\pm$0.1       & 29  $\pm$  3  \\
		SN2005cs   & 0.44 $\pm$ 0.03    &  4.0$\pm$0.1       & 24  $\pm$  3  \\

		SN1999em   & 0.31 $\pm$ 0.02    &  1.9$\pm$0.1       & 28  $\pm$  4  \\
 \hline
  \end{tabular}
\begin{flushleft}
Note: Objects are sorted in order of descending plateau slope.\\
  $^{a}$ Plateau slope during the linear decline phase, starting after first minima until plateau end.\\
  $^{b}$ Drop in magnitude during the plateau to nebular transition.\\
  $^{c}$ Duration of plateau to nebular transition.\\
  $^{i}$ Slope is calculated up to the available range of data, as plateau end is not observed.\\
\end{flushleft}
\end{table}

%% file: photstar.tex
\begin{table*}
 \centering
  \caption{Calibrated secondary standards in the fields of \sne\ with corresponding coordinates ($\alpha, \delta$) and calibrated magnitudes in \textit{UBVRI} bands are listed.  Errors quoted here include both photometric and calibration errors.}
  \label{tab:photstar}

  \begin{tabular}{cccccccc}
  \multicolumn{8}{c}{\snhj\ field standards}\\
     \hline
     Star& $\alpha_{\rm J2000}$&       $\delta_{\rm J2000}$& $U$& $B$&  $V$&  $R$&  $I$\\
       ID&          (h m s)&(\degr\,\arcmin\,\arcsec)&(mag)&(mag)&(mag)&(mag)&(mag)\\
     \hline
A   & 9:11:51.9 & -15:26:02.4 & 18.116 $\pm$ 0.071 & 17.272 $\pm$ 0.021 & 16.355 $\pm$ 0.012 & 15.824 $\pm$ 0.007 & 15.399 $\pm$ 0.024 \\
B   & 9:12:01.2 & -15:26:30.1 & 16.972 $\pm$ 0.032 & 17.011 $\pm$ 0.009 & 16.455 $\pm$ 0.010 & 16.136 $\pm$ 0.018 & 15.831 $\pm$ 0.022 \\
C   & 9:11:57.3 & -15:28:20.6 & 18.496 $\pm$ 0.114 & 18.896 $\pm$ 0.023 & 18.285 $\pm$ 0.021 & 17.934 $\pm$ 0.013 & 17.607 $\pm$ 0.030 \\
D   & 9:12:03.3 & -15:27:09.8 & 16.320 $\pm$ 0.013 & 16.414 $\pm$ 0.007 & 16.003 $\pm$ 0.009 & 15.742 $\pm$ 0.009 & 15.509 $\pm$ 0.017 \\
E   & 9:12:06.5 & -15:26:28.6 &          ---       & 19.144 $\pm$ 0.065 & 17.857 $\pm$ 0.021 & 17.056 $\pm$ 0.027 & 16.380 $\pm$ 0.022 \\
F   & 9:12:01.8 & -15:29:00.0 & 13.942 $\pm$ 0.010 & 13.615 $\pm$ 0.013 & 12.833 $\pm$ 0.015 & 12.311 $\pm$ 0.015 & 11.917 $\pm$ 0.062 \\
G   & 9:12:09.0 & -15:27:32.0 & 18.843 $\pm$ 0.110 & 18.105 $\pm$ 0.019 & 17.144 $\pm$ 0.009 & 16.551 $\pm$ 0.009 & 16.084 $\pm$ 0.034 \\
H   & 9:12:11.7 & -15:28:33.6 & 17.775 $\pm$ 0.065 & 16.867 $\pm$ 0.006 & 15.870 $\pm$ 0.010 & 15.308 $\pm$ 0.008 & 14.872 $\pm$ 0.018 \\
I   & 9:12:20.0 & -15:24:01.5 & 17.886 $\pm$ 0.085 & 17.959 $\pm$ 0.025 & 17.459 $\pm$ 0.023 & 17.090 $\pm$ 0.023 & 16.804 $\pm$ 0.019 \\
J   & 9:12:17.9 & -15:22:52.8 & 18.224 $\pm$ 0.095 & 17.938 $\pm$ 0.012 & 17.154 $\pm$ 0.015 & 16.727 $\pm$ 0.008 & 16.390 $\pm$ 0.017 \\
K   & 9:12:14.3 & -15:22:54.7 & 16.526 $\pm$ 0.017 & 16.024 $\pm$ 0.008 & 15.135 $\pm$ 0.010 & 14.620 $\pm$ 0.024 & 14.194 $\pm$ 0.014 \\
L   & 9:12:06.7 & -15:24:58.8 & 17.013 $\pm$ 0.027 & 17.041 $\pm$ 0.011 & 16.406 $\pm$ 0.012 & 16.043 $\pm$ 0.007 & 15.711 $\pm$ 0.021 \\
M   & 9:12:09.6 & -15:21:45.3 & 15.877 $\pm$ 0.010 & 15.860 $\pm$ 0.006 & 15.300 $\pm$ 0.012 & 14.944 $\pm$ 0.008 & 14.630 $\pm$ 0.013 \\
N   & 9:11:57.9 & -15:23:46.1 & 17.402 $\pm$ 0.028 & 17.060 $\pm$ 0.010 & 16.328 $\pm$ 0.014 & 15.901 $\pm$ 0.007 & 15.533 $\pm$ 0.029 \\
O   & 9:11:57.7 & -15:25:17.1 &          ---       & 18.942 $\pm$ 0.026 & 18.008 $\pm$ 0.027 & 17.399 $\pm$ 0.009 & 16.920 $\pm$ 0.023 \\
P   & 9:12:06.0 & -15:24:15.8 & 11.697 $\pm$ 0.006 & 11.660 $\pm$ 0.006 & 11.060 $\pm$ 0.031 &          ---       &          ---       \\
Q   & 9:12:08.6 & -15:26:57.0 & 12.162 $\pm$ 0.004 & 12.063 $\pm$ 0.005 & 11.796 $\pm$ 0.022 & 11.606 $\pm$ 0.015 & 11.442 $\pm$ 0.062 \\
\hline
\multicolumn{8}{c}{\sng\ field standards}\\
\hline
A   & 10:54:26.8 & 54:16:47.1 & 19.934 $\pm$ 0.456 & 18.943 $\pm$ 0.020 & 17.862 $\pm$ 0.020 & 17.238 $\pm$ 0.035 & 16.722 $\pm$ 0.042 \\
B   & 10:54:32.0 & 54:14:28.0 & 17.940 $\pm$ 0.044 & 17.854 $\pm$ 0.020 & 17.253 $\pm$ 0.016 & 16.880 $\pm$ 0.026 & 16.496 $\pm$ 0.030 \\
C   & 10:54:42.6 & 54:17:03.7 & 19.067 $\pm$ 0.080 & 18.648 $\pm$ 0.019 & 17.787 $\pm$ 0.023 & 17.284 $\pm$ 0.033 & 16.818 $\pm$ 0.041 \\
D   & 10:54:52.6 & 54:13:53.8 & 16.133 $\pm$ 0.022 & 16.106 $\pm$ 0.015 & 15.500 $\pm$ 0.018 & 15.144 $\pm$ 0.028 & 14.839 $\pm$ 0.031 \\
E   & 10:54:59.9 & 54:15:21.2 & 17.539 $\pm$ 0.024 & 17.626 $\pm$ 0.012 & 17.067 $\pm$ 0.014 & 16.714 $\pm$ 0.027 & 16.402 $\pm$ 0.039 \\
F   & 10:54:58.9 & 54:17:19.9 & 15.753 $\pm$ 0.009 & 15.141 $\pm$ 0.011 & 14.274 $\pm$ 0.016 & 13.754 $\pm$ 0.028 & 13.382 $\pm$ 0.039 \\
G   & 10:55:10.7 & 54:17:52.9 & 18.910 $\pm$ 0.097 & 17.656 $\pm$ 0.027 & 16.502 $\pm$ 0.017 & 15.796 $\pm$ 0.032 & 15.199 $\pm$ 0.025 \\
H   & 10:55:04.1 & 54:22:09.0 & 18.734 $\pm$ 0.107 & 18.595 $\pm$ 0.042 & 17.850 $\pm$ 0.019 & 17.445 $\pm$ 0.033 & 17.061 $\pm$ 0.023 \\
I   & 10:54:55.1 & 54:21:18.4 & 18.482 $\pm$ 0.062 & 18.552 $\pm$ 0.024 & 17.894 $\pm$ 0.020 & 17.533 $\pm$ 0.021 & 17.200 $\pm$ 0.023 \\
J   & 10:54:45.1 & 54:20:45.1 &          ---       & 19.994 $\pm$ 0.099 & 18.608 $\pm$ 0.031 & 17.746 $\pm$ 0.024 & 16.995 $\pm$ 0.024 \\
K   & 10:54:36.8 & 54:21:35.4 & 16.861 $\pm$ 0.013 & 16.252 $\pm$ 0.021 & 15.395 $\pm$ 0.017 & 14.899 $\pm$ 0.025 & 14.491 $\pm$ 0.024 \\
L   & 10:54:37.0 & 54:23:42.9 & 18.337 $\pm$ 0.060 & 18.915 $\pm$ 0.026 & 18.633 $\pm$ 0.045 & 18.191 $\pm$ 0.038 & 17.843 $\pm$ 0.038 \\
M   & 10:54:30.3 & 54:20:16.0 &          ---       & 19.328 $\pm$ 0.025 & 17.987 $\pm$ 0.026 & 17.215 $\pm$ 0.024 & 16.566 $\pm$ 0.035 \\
N   & 10:54:28.0 & 54:20:54.3 & 16.065 $\pm$ 0.018 & 16.137 $\pm$ 0.013 & 15.637 $\pm$ 0.016 & 15.309 $\pm$ 0.026 & 15.023 $\pm$ 0.028 \\
O   & 10:54:22.6 & 54:19:00.3 & 15.632 $\pm$ 0.014 & 15.407 $\pm$ 0.013 & 14.750 $\pm$ 0.014 & 14.362 $\pm$ 0.030 & 14.048 $\pm$ 0.034 \\
P   & 10:54:07.0 & 54:19:03.0 & 17.292 $\pm$ 0.024 & 17.089 $\pm$ 0.012 & 16.409 $\pm$ 0.023 & 15.976 $\pm$ 0.027 & 15.590 $\pm$ 0.032 \\
Q   & 10:54:01.6 & 54:18:04.5 & 18.828 $\pm$ 0.169 & 18.030 $\pm$ 0.035 & 17.049 $\pm$ 0.020 & 16.454 $\pm$ 0.030 & 15.865 $\pm$ 0.034 \\

     \hline
  \end{tabular}

\end{table*}

%% file: photsnhj.tex
\onecolumn
{\centering
 \fontsize{2.0mm}{2.6mm}\selectfont
  \begin{longtable}
  {c c r c c c c c l c}
  \caption{Photometric evolution of \snhj\ in \textit{UBVRI} bands.}
  \label{tab:photsnhj}\\
  \hline
  UT Date&JD&Phase$^{a}$&$U$&$B$&$V$&$R$&$I$&Tel$^{b}$ \\
  (yyyy-mm-dd)&2456000+&(day)&(mag)&(mag)&(mag)&(mag)&(mag)& \\
  \hline
     \endfirsthead
     \multicolumn{3}{l}{ {\tablename\ \thetable\ - continued. }} \\
     \hline
  UT Date&JD&Phase$^{a}$&$U$&$B$&$V$&$R$&$I$&Tel$^{b}$ \\
    (yyyy-mm-dd)&2456000+&(day)&(mag)&(mag)&(mag)&(mag)&(mag)& \\
     \hline
     \endhead
     \hline
     \endfoot
     \hline
     \endlastfoot
2013-12-13.88  &   640.38   &    3.38   &   13.681 $\pm$ 0.041 &  14.700 $\pm$ 0.024 &  14.819 $\pm$ 0.016 &  14.806 $\pm$ 0.016 &  14.858 $\pm$ 0.034 &     ARIES  \\
2013-12-14.84  &   641.34   &    4.34   &   13.721 $\pm$ 0.025 &  14.702 $\pm$ 0.014 &  14.757 $\pm$ 0.010 &  14.716 $\pm$ 0.010 &  14.743 $\pm$ 0.020 &     ARIES  \\
2013-12-15.92  &   642.42   &    5.42   &   13.739 $\pm$ 0.028 &  14.699 $\pm$ 0.017 &  14.696 $\pm$ 0.011 &  14.633 $\pm$ 0.011 &  14.639 $\pm$ 0.023 &     ARIES  \\
2013-12-16.93  &   643.43   &    6.43   &   13.775 $\pm$ 0.028 &  14.694 $\pm$ 0.024 &  14.667 $\pm$ 0.016 &  14.565 $\pm$ 0.015 &  14.546 $\pm$ 0.031 &     ARIES  \\
2013-12-18.79  &   645.29   &    8.29   &   13.849 $\pm$ 0.042 &  14.738 $\pm$ 0.024 &  14.632 $\pm$ 0.016 &  14.492 $\pm$ 0.015 &  14.441 $\pm$ 0.031 &     ARIES  \\
2013-12-19.80  &   646.30   &    9.30   &   13.899 $\pm$ 0.030 &  14.769 $\pm$ 0.018 &  14.653 $\pm$ 0.012 &  14.496 $\pm$ 0.011 &  14.412 $\pm$ 0.021 &     ARIES  \\
2013-12-21.01  &   647.51   &   10.51   &   13.924 $\pm$ 0.047 &  14.758 $\pm$ 0.026 &  14.652 $\pm$ 0.018 &  14.505 $\pm$ 0.016 &  14.388 $\pm$ 0.034 &     ARIES  \\
2013-12-25.83  &   652.33   &   15.33   &   14.151 $\pm$ 0.021 &  14.892 $\pm$ 0.017 &  14.730 $\pm$ 0.010 &  14.496 $\pm$ 0.011 &  14.412 $\pm$ 0.023 &     ARIES  \\
2013-12-28.72  &   655.22   &   18.22   &           ---        &          ---        &  14.785 $\pm$ 0.012 &  14.558 $\pm$ 0.013 &          ---        &       OKU  \\
2013-12-28.83  &   655.33   &   18.33   &   14.281 $\pm$ 0.029 &  14.943 $\pm$ 0.017 &  14.738 $\pm$ 0.012 &  14.516 $\pm$ 0.011 &  14.447 $\pm$ 0.023 &     ARIES  \\
2013-12-29.73  &   656.23   &   19.23   &           ---        &          ---        &  14.783 $\pm$ 0.008 &  14.579 $\pm$ 0.008 &          ---        &       OKU  \\
2013-12-29.91  &   656.41   &   19.41   &   14.371 $\pm$ 0.029 &  14.969 $\pm$ 0.017 &  14.753 $\pm$ 0.011 &          ---        &  14.467 $\pm$ 0.023 &     ARIES  \\
2013-12-31.69  &   658.19   &   21.19   &           ---        &          ---        &          ---        &          ---        &  14.490 $\pm$ 0.033 &       OKU  \\
2014-01-01.66  &   659.16   &   22.16   &           ---        &          ---        &  14.835 $\pm$ 0.008 &  14.582 $\pm$ 0.008 &          ---        &       OKU  \\
2014-01-02.69  &   660.19   &   23.19   &           ---        &          ---        &  14.829 $\pm$ 0.008 &  14.598 $\pm$ 0.009 &          ---        &       OKU  \\
2014-01-03.68  &   661.18   &   24.18   &           ---        &          ---        &  14.848 $\pm$ 0.008 &  14.608 $\pm$ 0.009 &          ---        &       OKU  \\
2014-01-05.61  &   663.11   &   26.11   &           ---        &          ---        &  14.849 $\pm$ 0.009 &  14.576 $\pm$ 0.007 &          ---        &       OKU  \\
2014-01-06.61  &   664.11   &   27.11   &           ---        &  15.304 $\pm$ 0.015 &          ---        &          ---        &  14.487 $\pm$ 0.015 &       OKU  \\
2014-01-06.86  &   664.36   &   27.36   &           ---        &  15.280 $\pm$ 0.018 &  14.919 $\pm$ 0.012 &  14.606 $\pm$ 0.011 &  14.515 $\pm$ 0.023 &     ARIES  \\
2014-01-07.73  &   665.23   &   28.23   &           ---        &          ---        &  14.890 $\pm$ 0.014 &          ---        &  14.481 $\pm$ 0.022 &       OKU  \\
2014-01-09.72  &   667.22   &   30.22   &           ---        &          ---        &          ---        &  14.587 $\pm$ 0.010 &  14.489 $\pm$ 0.026 &       OKU  \\
2014-01-10.56  &   668.06   &   31.06   &           ---        &          ---        &  14.928 $\pm$ 0.010 &  14.632 $\pm$ 0.009 &          ---        &       OKU  \\
2014-01-11.59  &   669.09   &   32.09   &           ---        &          ---        &          ---        &          ---        &  14.491 $\pm$ 0.012 &       OKU  \\
2014-01-11.97  &   669.47   &   32.47   &   15.610 $\pm$ 0.037 &  15.519 $\pm$ 0.026 &  14.969 $\pm$ 0.016 &  14.622 $\pm$ 0.015 &  14.527 $\pm$ 0.032 &     ARIES  \\
2014-01-12.79  &   670.29   &   33.29   &   15.695 $\pm$ 0.037 &  15.568 $\pm$ 0.020 &  14.969 $\pm$ 0.012 &  14.626 $\pm$ 0.009 &  14.516 $\pm$ 0.023 &     ARIES  \\
2014-01-13.66  &   671.16   &   34.16   &           ---        &          ---        &  14.981 $\pm$ 0.007 &          ---        &          ---        &       OKU  \\
2014-01-14.80  &   672.30   &   35.30   &           ---        &          ---        &          ---        &  14.633 $\pm$ 0.015 &  14.533 $\pm$ 0.031 &     ARIES  \\
2014-01-15.64  &   673.14   &   36.14   &           ---        &  15.700 $\pm$ 0.017 &  14.989 $\pm$ 0.009 &          ---        &          ---        &       OKU  \\
2014-01-15.93  &   673.43   &   36.43   &   15.854 $\pm$ 0.048 &          ---        &  14.995 $\pm$ 0.017 &  14.642 $\pm$ 0.015 &  14.547 $\pm$ 0.032 &     ARIES  \\
2014-01-16.71  &   674.21   &   37.21   &           ---        &  15.782 $\pm$ 0.039 &          ---        &  14.650 $\pm$ 0.009 &  14.503 $\pm$ 0.020 &       OKU  \\
2014-01-16.82  &   674.32   &   37.32   &   16.073 $\pm$ 0.042 &  15.744 $\pm$ 0.021 &  15.017 $\pm$ 0.012 &  14.649 $\pm$ 0.011 &  14.541 $\pm$ 0.023 &     ARIES  \\
2014-01-17.71  &   675.21   &   38.21   &           ---        &          ---        &  15.038 $\pm$ 0.011 &  14.663 $\pm$ 0.011 &          ---        &       OKU  \\
2014-01-18.83  &   676.33   &   39.33   &           ---        &          ---        &          ---        &          ---        &  14.504 $\pm$ 0.021 &       OKU  \\
2014-01-19.71  &   677.21   &   40.21   &           ---        &  15.833 $\pm$ 0.021 &  15.077 $\pm$ 0.011 &  14.691 $\pm$ 0.011 &          ---        &       OKU  \\
2014-01-20.85  &   678.35   &   41.35   &   16.418 $\pm$ 0.051 &  15.879 $\pm$ 0.019 &  15.092 $\pm$ 0.010 &  14.687 $\pm$ 0.010 &  14.517 $\pm$ 0.015 &     ARIES  \\
2014-01-21.71  &   679.21   &   42.21   &           ---        &          ---        &  15.116 $\pm$ 0.021 &  14.706 $\pm$ 0.038 &  14.532 $\pm$ 0.021 &       OKU  \\
2014-01-23.75  &   681.25   &   44.25   &           ---        &  15.995 $\pm$ 0.020 &  15.155 $\pm$ 0.010 &          ---        &          ---        &       OKU  \\
2014-01-24.80  &   682.30   &   45.30   &           ---        &          ---        &          ---        &          ---        &  14.590 $\pm$ 0.017 &       OKU  \\
2014-01-24.84  &   682.34   &   45.34   &   16.774 $\pm$ 0.047 &  15.963 $\pm$ 0.022 &  15.237 $\pm$ 0.013 &  14.783 $\pm$ 0.010 &  14.568 $\pm$ 0.023 &     ARIES  \\
2014-01-25.83  &   683.33   &   46.33   &   16.627 $\pm$ 0.087 &  15.982 $\pm$ 0.029 &  15.148 $\pm$ 0.016 &  14.683 $\pm$ 0.015 &  14.566 $\pm$ 0.031 &     ARIES  \\
2014-01-26.82  &   684.32   &   47.32   &   16.687 $\pm$ 0.061 &  16.018 $\pm$ 0.028 &  15.168 $\pm$ 0.016 &  14.694 $\pm$ 0.015 &  14.577 $\pm$ 0.031 &     ARIES  \\
2014-01-27.74  &   685.24   &   48.24   &           ---        &  16.095 $\pm$ 0.022 &  15.202 $\pm$ 0.010 &          ---        &          ---        &       OKU  \\
2014-01-28.75  &   686.25   &   49.25   &           ---        &          ---        &          ---        &  14.739 $\pm$ 0.009 &  14.579 $\pm$ 0.017 &       OKU  \\
2014-01-28.90  &   686.40   &   49.40   &   16.890 $\pm$ 0.098 &  16.157 $\pm$ 0.024 &  15.207 $\pm$ 0.013 &  14.742 $\pm$ 0.013 &  14.611 $\pm$ 0.020 &     ARIES  \\
2014-01-29.84  &   687.34   &   50.34   &   16.879 $\pm$ 0.068 &  16.118 $\pm$ 0.030 &  15.210 $\pm$ 0.016 &  14.720 $\pm$ 0.015 &  14.596 $\pm$ 0.031 &     ARIES  \\
2014-01-30.82  &   688.32   &   51.32   &   17.009 $\pm$ 0.065 &  16.215 $\pm$ 0.025 &  15.247 $\pm$ 0.011 &  14.775 $\pm$ 0.012 &  14.633 $\pm$ 0.023 &     ARIES  \\
2014-01-31.73  &   689.23   &   52.23   &           ---        &  16.324 $\pm$ 0.033 &          ---        &          ---        &          ---        &       OKU  \\
2014-01-31.80  &   689.30   &   52.30   &   17.049 $\pm$ 0.064 &  16.266 $\pm$ 0.031 &  15.270 $\pm$ 0.016 &  14.778 $\pm$ 0.015 &  14.641 $\pm$ 0.031 &     ARIES  \\
2014-02-01.78  &   690.28   &   53.28   &   17.093 $\pm$ 0.064 &  16.279 $\pm$ 0.026 &  15.276 $\pm$ 0.013 &  14.788 $\pm$ 0.012 &  14.642 $\pm$ 0.023 &     ARIES  \\
2014-02-02.76  &   691.26   &   54.26   &   17.152 $\pm$ 0.070 &  16.322 $\pm$ 0.033 &  15.300 $\pm$ 0.016 &  14.798 $\pm$ 0.015 &  14.645 $\pm$ 0.031 &     ARIES  \\
2014-02-05.76  &   694.26   &   57.26   &           ---        &  16.310 $\pm$ 0.027 &  15.323 $\pm$ 0.012 &          ---        &          ---        &       OKU  \\
2014-02-05.84  &   694.34   &   57.34   &   17.300 $\pm$ 0.089 &  16.306 $\pm$ 0.032 &  15.313 $\pm$ 0.017 &  14.776 $\pm$ 0.015 &  14.628 $\pm$ 0.032 &     ARIES  \\
2014-02-09.71  &   698.21   &   61.21   &           ---        &          ---        &  15.370 $\pm$ 0.013 &  14.826 $\pm$ 0.035 &          ---        &       OKU  \\
2014-02-09.79  &   698.29   &   61.29   &   17.510 $\pm$ 0.104 &  16.510 $\pm$ 0.036 &  15.403 $\pm$ 0.013 &  14.868 $\pm$ 0.016 &  14.722 $\pm$ 0.031 &     ARIES  \\
2014-02-10.69  &   699.19   &   62.19   &           ---        &          ---        &          ---        &  14.859 $\pm$ 0.009 &  14.682 $\pm$ 0.019 &       OKU  \\
2014-02-11.82  &   700.32   &   63.32   &   17.701 $\pm$ 0.118 &  16.573 $\pm$ 0.030 &  15.430 $\pm$ 0.011 &  14.895 $\pm$ 0.012 &  14.734 $\pm$ 0.024 &     ARIES  \\
2014-02-12.87  &   701.37   &   64.37   &   17.708 $\pm$ 0.144 &  16.543 $\pm$ 0.031 &  15.439 $\pm$ 0.011 &  14.896 $\pm$ 0.010 &  14.748 $\pm$ 0.020 &     ARIES  \\
2014-02-16.75  &   705.25   &   68.25   &   17.619 $\pm$ 0.120 &  16.664 $\pm$ 0.035 &  15.502 $\pm$ 0.014 &  14.941 $\pm$ 0.012 &  14.779 $\pm$ 0.024 &     ARIES  \\
2014-02-17.74  &   706.24   &   69.24   &   17.894 $\pm$ 0.132 &  16.728 $\pm$ 0.035 &  15.515 $\pm$ 0.013 &  14.941 $\pm$ 0.012 &  14.764 $\pm$ 0.024 &     ARIES  \\
2014-02-19.57  &   708.07   &   71.07   &           ---        &          ---        &          ---        &  15.006 $\pm$ 0.013 &          ---        &       OKU  \\
2014-02-19.86  &   708.36   &   71.36   &   17.910 $\pm$ 0.173 &  16.757 $\pm$ 0.037 &  15.567 $\pm$ 0.018 &  14.975 $\pm$ 0.016 &  14.807 $\pm$ 0.033 &     ARIES  \\
2014-02-20.72  &   709.22   &   72.22   &           ---        &          ---        &          ---        &  14.975 $\pm$ 0.010 &  14.813 $\pm$ 0.019 &       OKU  \\
2014-02-21.69  &   710.19   &   73.19   &   18.064 $\pm$ 0.138 &  16.705 $\pm$ 0.032 &  15.607 $\pm$ 0.018 &  14.967 $\pm$ 0.016 &  14.798 $\pm$ 0.024 &     ARIES  \\
2014-02-22.75  &   711.25   &   74.25   &   18.163 $\pm$ 0.156 &  16.754 $\pm$ 0.040 &  15.619 $\pm$ 0.018 &  14.976 $\pm$ 0.016 &  14.808 $\pm$ 0.024 &     ARIES  \\
2014-02-23.75  &   712.25   &   75.25   &           ---        &          ---        &          ---        &          ---        &          ---        &       OKU  \\
2014-02-24.61  &   713.11   &   76.11   &           ---        &  16.793 $\pm$ 0.039 &  15.619 $\pm$ 0.013 &          ---        &          ---        &       OKU  \\
2014-02-25.64  &   714.14   &   77.14   &           ---        &          ---        &  15.627 $\pm$ 0.018 &  15.059 $\pm$ 0.015 &          ---        &       OKU  \\
2014-02-25.80  &   714.30   &   77.30   &   18.080 $\pm$ 0.181 &  16.901 $\pm$ 0.050 &  15.608 $\pm$ 0.018 &  15.047 $\pm$ 0.013 &  14.862 $\pm$ 0.024 &     ARIES  \\
2014-03-02.70  &   719.20   &   82.20   &   18.537 $\pm$ 0.220 &  17.071 $\pm$ 0.045 &  15.732 $\pm$ 0.015 &  15.147 $\pm$ 0.013 &  14.943 $\pm$ 0.024 &     ARIES  \\
2014-03-03.62  &   720.12   &   83.12   &           ---        &          ---        &  15.851 $\pm$ 0.016 &          ---        &          ---        &       OKU  \\
2014-03-05.69  &   722.19   &   85.19   &           ---        &          ---        &          ---        &          ---        &  14.985 $\pm$ 0.054 &       OKU  \\
2014-03-07.68  &   724.18   &   87.18   &           ---        &          ---        &  15.907 $\pm$ 0.019 &  15.274 $\pm$ 0.019 &          ---        &       OKU  \\
2014-03-07.69  &   724.19   &   87.19   &   18.786 $\pm$ 0.266 &  17.292 $\pm$ 0.053 &  15.910 $\pm$ 0.016 &  15.229 $\pm$ 0.013 &  15.037 $\pm$ 0.025 &     ARIES  \\
2014-03-08.73  &   725.23   &   88.23   &   18.852 $\pm$ 0.293 &  17.336 $\pm$ 0.052 &  15.942 $\pm$ 0.017 &  15.262 $\pm$ 0.013 &  15.064 $\pm$ 0.034 &     ARIES  \\
2014-03-10.55  &   727.05   &   90.05   &           ---        &          ---        &  15.995 $\pm$ 0.020 &  15.311 $\pm$ 0.012 &          ---        &       OKU  \\
2014-03-11.67  &   728.17   &   91.17   &           ---        &          ---        &  16.003 $\pm$ 0.025 &  15.358 $\pm$ 0.019 &          ---        &       OKU  \\
2014-03-12.76  &   729.26   &   92.26   &           ---        &  17.478 $\pm$ 0.065 &  16.065 $\pm$ 0.022 &  15.365 $\pm$ 0.018 &  15.143 $\pm$ 0.027 &     ARIES  \\
2014-03-13.68  &   730.18   &   93.18   &   18.927 $\pm$ 0.393 &  17.565 $\pm$ 0.069 &  16.095 $\pm$ 0.018 &  15.381 $\pm$ 0.015 &  15.165 $\pm$ 0.026 &     ARIES  \\
2014-03-14.62  &   731.12   &   94.12   &           ---        &          ---        &  16.187 $\pm$ 0.019 &  15.388 $\pm$ 0.023 &          ---        &       OKU  \\
2014-03-14.83  &   731.33   &   94.33   &           ---        &  17.561 $\pm$ 0.076 &  16.116 $\pm$ 0.019 &  15.434 $\pm$ 0.013 &  15.204 $\pm$ 0.023 &     ARIES  \\
2014-03-16.80  &   733.30   &   96.30   &           ---        &  17.740 $\pm$ 0.094 &  16.209 $\pm$ 0.020 &  15.460 $\pm$ 0.014 &  15.254 $\pm$ 0.028 &     ARIES  \\
2014-03-18.67  &   735.17   &   98.17   &           ---        &  17.787 $\pm$ 0.086 &  16.244 $\pm$ 0.024 &  15.551 $\pm$ 0.018 &  15.313 $\pm$ 0.028 &     ARIES  \\
2014-03-19.67  &   736.17   &   99.17   &   18.781 $\pm$ 0.317 &  17.853 $\pm$ 0.088 &  16.319 $\pm$ 0.021 &  15.572 $\pm$ 0.017 &  15.351 $\pm$ 0.028 &     ARIES  \\
2014-03-20.67  &   737.17   &  100.17   &   19.102 $\pm$ 0.361 &  17.909 $\pm$ 0.089 &  16.388 $\pm$ 0.021 &  15.620 $\pm$ 0.015 &  15.390 $\pm$ 0.025 &     ARIES  \\
2014-03-23.62  &   740.12   &  103.12   &           ---        &          ---        &          ---        &  15.827 $\pm$ 0.021 &  15.532 $\pm$ 0.031 &       OKU  \\
2014-03-23.79  &   740.29   &  103.29   &           ---        &  18.099 $\pm$ 0.107 &  16.564 $\pm$ 0.030 &  15.761 $\pm$ 0.022 &  15.550 $\pm$ 0.039 &     ARIES  \\
2014-03-27.63  &   744.13   &  107.13   &           ---        &          ---        &          ---        &  16.095 $\pm$ 0.030 &  15.811 $\pm$ 0.033 &       OKU  \\
2014-03-31.63  &   748.13   &  111.13   &           ---        &          ---        &          ---        &  16.357 $\pm$ 0.030 &  16.250 $\pm$ 0.046 &       OKU  \\
2014-04-01.61  &   749.11   &  112.11   &           ---        &          ---        &          ---        &  16.448 $\pm$ 0.050 &          ---        &       OKU  \\
2014-04-02.59  &   750.09   &  113.09   &           ---        &  18.897 $\pm$ 0.217 &  17.533 $\pm$ 0.066 &  16.581 $\pm$ 0.038 &  16.324 $\pm$ 0.057 &     ARIES  \\
2014-04-06.60  &   754.10   &  117.10   &           ---        &          ---        &          ---        &  16.672 $\pm$ 0.038 &          ---        &       OKU  \\
2014-04-07.62  &   755.12   &  118.12   &           ---        &          ---        &          ---        &          ---        &  16.387 $\pm$ 0.053 &       OKU  \\
2014-04-07.73  &   755.23   &  118.23   &           ---        &          ---        &  17.699 $\pm$ 0.070 &  16.723 $\pm$ 0.041 &  16.438 $\pm$ 0.054 &     ARIES  \\
2014-04-08.57  &   756.07   &  119.07   &           ---        &          ---        &          ---        &  16.737 $\pm$ 0.046 &          ---        &       OKU  \\
2014-04-08.63  &   756.13   &  119.13   &           ---        &  19.141 $\pm$ 0.268 &  17.809 $\pm$ 0.071 &  16.789 $\pm$ 0.039 &  16.516 $\pm$ 0.056 &     ARIES  \\
2014-04-09.57  &   757.07   &  120.07   &           ---        &          ---        &          ---        &  16.829 $\pm$ 0.047 &          ---        &       OKU  \\
2014-04-09.62  &   757.12   &  120.12   &           ---        &  19.218 $\pm$ 0.303 &  17.866 $\pm$ 0.081 &  16.831 $\pm$ 0.044 &  16.555 $\pm$ 0.061 &     ARIES  \\
2014-04-11.47  &   758.97   &  121.97   &           ---        &          ---        &          ---        &  16.833 $\pm$ 0.052 &          ---        &       OKU  \\
2014-04-11.72  &   759.22   &  122.22   &           ---        &  18.986 $\pm$ 0.289 &  17.722 $\pm$ 0.079 &  16.765 $\pm$ 0.039 &  16.522 $\pm$ 0.061 &     ARIES  \\
2014-04-12.56  &   760.06   &  123.06   &           ---        &          ---        &          ---        &          ---        &          ---        &       OKU  \\
2014-04-14.65  &   762.15   &  125.15   &           ---        &  19.470 $\pm$ 0.409 &  17.755 $\pm$ 0.070 &  16.842 $\pm$ 0.043 &  16.563 $\pm$ 0.055 &     ARIES  \\
2014-04-19.68  &   767.18   &  130.18   &   19.739 $\pm$ 0.707 &  19.592 $\pm$ 0.418 &  17.847 $\pm$ 0.071 &  16.875 $\pm$ 0.044 &  16.612 $\pm$ 0.061 &     ARIES  \\
2014-04-20.65  &   768.15   &  131.15   &           ---        &  19.274 $\pm$ 0.301 &  17.958 $\pm$ 0.080 &  16.897 $\pm$ 0.043 &  16.636 $\pm$ 0.063 &     ARIES  \\
2014-04-21.64  &   769.14   &  132.14   &   20.197 $\pm$ 0.955 &  19.281 $\pm$ 0.307 &  17.920 $\pm$ 0.077 &  16.894 $\pm$ 0.049 &  16.638 $\pm$ 0.061 &     ARIES  \\
2014-04-23.55  &   771.05   &  134.05   &           ---        &          ---        &          ---        &          ---        &  16.617 $\pm$ 0.063 &       OKU  \\
2014-04-26.74  &   774.24   &  137.24   &           ---        &  19.490 $\pm$ 0.477 &  17.986 $\pm$ 0.112 &  16.910 $\pm$ 0.047 &  16.619 $\pm$ 0.063 &     ARIES  \\
2014-04-27.63  &   775.13   &  138.13   &   19.730 $\pm$ 0.809 &  19.457 $\pm$ 0.390 &  17.906 $\pm$ 0.089 &  16.941 $\pm$ 0.046 &  16.635 $\pm$ 0.063 &     ARIES  \\
2014-05-09.63  &   787.13   &  150.13   &           ---        &  19.409 $\pm$ 0.473 &  18.021 $\pm$ 0.095 &  17.003 $\pm$ 0.055 &  16.693 $\pm$ 0.064 &     ARIES  \\
2014-05-10.63  &   788.13   &  151.13   &           ---        &          ---        &  18.063 $\pm$ 0.093 &  17.061 $\pm$ 0.052 &  16.724 $\pm$ 0.073 &     ARIES  \\
2014-05-13.63  &   791.13   &  154.13   &           ---        &  19.081 $\pm$ 0.401 &  17.975 $\pm$ 0.086 &  17.096 $\pm$ 0.061 &  16.917 $\pm$ 0.089 &     ARIES  \\
2014-05-15.64  &   793.14   &  156.14   &           ---        &  19.276 $\pm$ 0.355 &  17.947 $\pm$ 0.089 &  17.128 $\pm$ 0.061 &  16.709 $\pm$ 0.072 &     ARIES  \\
2014-05-16.62  &   794.12   &  157.12   &           ---        &  19.408 $\pm$ 0.366 &  17.923 $\pm$ 0.087 &  17.093 $\pm$ 0.057 &  16.759 $\pm$ 0.072 &     ARIES  \\
2014-05-20.62  &   798.12   &  161.12   &           ---        &  19.469 $\pm$ 0.362 &  18.233 $\pm$ 0.104 &  17.165 $\pm$ 0.056 &  16.894 $\pm$ 0.076 &     ARIES  \\
2014-05-21.63  &   799.13   &  162.13   &           ---        &  19.436 $\pm$ 0.361 &  18.253 $\pm$ 0.108 &  17.165 $\pm$ 0.056 &  16.868 $\pm$ 0.076 &     ARIES  \\
2014-05-26.63  &   804.13   &  167.13   &           ---        &          ---        &  18.075 $\pm$ 0.098 &  17.103 $\pm$ 0.060 &  16.818 $\pm$ 0.079 &     ARIES  \\
2014-05-29.63  &   807.13   &  170.13   &           ---        &          ---        &  18.138 $\pm$ 0.095 &  17.165 $\pm$ 0.061 &  16.939 $\pm$ 0.086 &     ARIES  \\
\hline

  \end{longtable}

\begin{flushleft}
  $^{a}$ with reference to the explosion epoch JD=\thj d\\
  $^{b}$ ARIES: 104cm Sampurnanand telescope and 130cm Devasthal fast optical telescope at ARIES, Nainital, India; OKU: 51cm telescope at Osaka Kyoiku University, Japan\\
  Note: Data observed within 5 Hrs, are represented under single epoch observation.
\end{flushleft}
}
\twocolumn

%% file: photsng.tex
\begin{table*}
\centering
  \fontsize{2.0mm}{2.5mm}\selectfont
  \caption{Photometric evolution of \sng.}
  \label{tab:photsng}
  \textit{UBVRI} photometry\\
  \begin{tabular}
  {c c r c c c c c l c}
  \hline
  UT Date&JD&Phase$^{a}$&$U$&$B$&$V$&$R$&$I$&Tel$^{b}$ \\
  (yyyy-mm-dd)&2456000+&(day)&(mag)&(mag)&(mag)&(mag)&(mag)& \\
  \hline
2014-01-15.97  &   673.47   &    3.74     & 14.215 $\pm$ 0.040  & 15.043 $\pm$ 0.023  & 15.007 $\pm$ 0.017  & 14.924 $\pm$ 0.017  & 14.900 $\pm$ 0.033  &  ARIES  \\
2014-01-16.88  &   674.38   &    4.65     & 14.085 $\pm$ 0.028  & 14.911 $\pm$ 0.016  & 14.851 $\pm$ 0.013  & 14.738 $\pm$ 0.012  & 14.693 $\pm$ 0.020  &  ARIES  \\
2014-01-20.92  &   678.42   &    8.69     & 14.110 $\pm$ 0.029  & 14.829 $\pm$ 0.016  & 14.601 $\pm$ 0.012  & 14.436 $\pm$ 0.011  & 14.311 $\pm$ 0.022  &  ARIES  \\
2014-01-24.92  &   682.42   &   12.69     & 14.310 $\pm$ 0.028  & 14.860 $\pm$ 0.016  & 14.451 $\pm$ 0.011  &          ---        & 14.064 $\pm$ 0.022  &  ARIES  \\
2014-01-25.93  &   683.43   &   13.70     & 14.294 $\pm$ 0.019  & 14.867 $\pm$ 0.014  & 14.472 $\pm$ 0.010  & 14.213 $\pm$ 0.012  & 14.067 $\pm$ 0.020  &  ARIES  \\
2014-01-27.91  &   685.41   &   15.68     & 14.441 $\pm$ 0.040  & 14.916 $\pm$ 0.023  & 14.465 $\pm$ 0.016  &          ---        & 13.997 $\pm$ 0.030  &  ARIES  \\
2014-01-28.89  &   686.39   &   16.66     & 14.534 $\pm$ 0.030  & 14.996 $\pm$ 0.016  & 14.561 $\pm$ 0.012  & 14.263 $\pm$ 0.011  & 14.023 $\pm$ 0.022  &  ARIES  \\
2014-01-29.89  &   687.39   &   17.66     & 14.599 $\pm$ 0.040  & 15.007 $\pm$ 0.023  & 14.516 $\pm$ 0.016  & 14.206 $\pm$ 0.015  & 14.006 $\pm$ 0.030  &  ARIES  \\
2014-01-30.91  &   688.41   &   18.68     & 14.689 $\pm$ 0.029  & 15.073 $\pm$ 0.016  & 14.609 $\pm$ 0.011  & 14.296 $\pm$ 0.011  & 14.044 $\pm$ 0.022  &  ARIES  \\
2014-01-31.83  &   689.33   &   19.60     & 14.780 $\pm$ 0.029  & 15.143 $\pm$ 0.023  & 14.644 $\pm$ 0.016  & 14.308 $\pm$ 0.015  & 14.052 $\pm$ 0.030  &  ARIES  \\
2014-02-01.84  &   690.34   &   20.61     & 14.859 $\pm$ 0.023  & 15.185 $\pm$ 0.016  & 14.676 $\pm$ 0.011  & 14.320 $\pm$ 0.011  & 14.062 $\pm$ 0.022  &  ARIES  \\
2014-02-02.86  &   691.36   &   21.63     & 14.954 $\pm$ 0.025  & 15.245 $\pm$ 0.019  & 14.686 $\pm$ 0.014  & 14.344 $\pm$ 0.011  & 14.084 $\pm$ 0.022  &  ARIES  \\
2014-02-08.94  &   697.44   &   27.71     &          ---        &          ---        &          ---        &          ---        &          ---        &  ARIES  \\
2014-02-09.82  &   698.32   &   28.59     & 15.662 $\pm$ 0.032  & 15.658 $\pm$ 0.017  & 14.962 $\pm$ 0.013  & 14.555 $\pm$ 0.012  & 14.270 $\pm$ 0.022  &  ARIES  \\
2014-02-11.88  &   700.38   &   30.65     & 15.890 $\pm$ 0.049  & 15.816 $\pm$ 0.017  & 15.051 $\pm$ 0.013  & 14.624 $\pm$ 0.012  & 14.331 $\pm$ 0.022  &  ARIES  \\
2014-02-12.91  &   701.41   &   31.68     & 16.013 $\pm$ 0.034  & 15.860 $\pm$ 0.024  & 15.064 $\pm$ 0.017  & 14.636 $\pm$ 0.015  & 14.333 $\pm$ 0.030  &  ARIES  \\
2014-02-16.81  &   705.31   &   35.58     & 16.467 $\pm$ 0.045  & 16.087 $\pm$ 0.020  & 15.231 $\pm$ 0.014  & 14.751 $\pm$ 0.013  & 14.447 $\pm$ 0.023  &  ARIES  \\
2014-02-17.80  &   706.30   &   36.57     & 16.516 $\pm$ 0.034  & 16.138 $\pm$ 0.019  & 15.232 $\pm$ 0.014  & 14.728 $\pm$ 0.012  & 14.423 $\pm$ 0.023  &  ARIES  \\
2014-02-19.89  &   708.39   &   38.66     & 16.814 $\pm$ 0.046  & 16.263 $\pm$ 0.019  & 15.314 $\pm$ 0.019  & 14.818 $\pm$ 0.017  & 14.491 $\pm$ 0.031  &  ARIES  \\
2014-02-21.74  &   710.24   &   40.51     & 16.910 $\pm$ 0.040  & 16.287 $\pm$ 0.020  & 15.365 $\pm$ 0.020  & 14.798 $\pm$ 0.017  & 14.498 $\pm$ 0.031  &  ARIES  \\
2014-02-22.78  &   711.28   &   41.55     & 16.968 $\pm$ 0.042  & 16.331 $\pm$ 0.027  & 15.397 $\pm$ 0.019  & 14.817 $\pm$ 0.016  & 14.512 $\pm$ 0.031  &  ARIES  \\
2014-02-25.84  &   714.34   &   44.61     & 17.399 $\pm$ 0.065  & 16.565 $\pm$ 0.022  & 15.469 $\pm$ 0.020  & 14.955 $\pm$ 0.017  & 14.601 $\pm$ 0.032  &  ARIES  \\
2014-03-02.77  &   719.27   &   49.54     & 17.705 $\pm$ 0.060  & 16.722 $\pm$ 0.023  & 15.616 $\pm$ 0.017  & 15.078 $\pm$ 0.014  & 14.710 $\pm$ 0.023  &  ARIES  \\
2014-03-07.76  &   724.26   &   54.53     & 17.957 $\pm$ 0.063  & 16.940 $\pm$ 0.025  & 15.768 $\pm$ 0.019  & 15.151 $\pm$ 0.015  & 14.804 $\pm$ 0.024  &  ARIES  \\
2014-03-08.77  &   725.27   &   55.54     & 18.019 $\pm$ 0.075  & 16.985 $\pm$ 0.024  & 15.784 $\pm$ 0.022  & 15.173 $\pm$ 0.018  & 14.816 $\pm$ 0.032  &  ARIES  \\
2014-03-12.79  &   729.29   &   59.56     & 18.301 $\pm$ 0.113  & 17.112 $\pm$ 0.036  & 15.879 $\pm$ 0.024  & 15.193 $\pm$ 0.018  & 14.884 $\pm$ 0.024  &  ARIES  \\
2014-03-13.75  &   730.25   &   60.52     & 18.428 $\pm$ 0.113  & 17.160 $\pm$ 0.027  & 15.892 $\pm$ 0.021  & 15.247 $\pm$ 0.015  & 14.887 $\pm$ 0.024  &  ARIES  \\
2014-03-14.87  &   731.37   &   61.64     & 18.596 $\pm$ 0.169  & 17.175 $\pm$ 0.039  & 15.914 $\pm$ 0.024  & 15.266 $\pm$ 0.019  & 14.916 $\pm$ 0.033  &  ARIES  \\
2014-03-16.83  &   733.33   &   63.60     & 18.276 $\pm$ 0.167  & 17.243 $\pm$ 0.031  & 15.957 $\pm$ 0.026  & 15.294 $\pm$ 0.020  & 14.931 $\pm$ 0.033  &  ARIES  \\
2014-03-18.72  &   735.22   &   65.49     & 18.426 $\pm$ 0.142  & 17.287 $\pm$ 0.033  & 15.997 $\pm$ 0.022  & 15.321 $\pm$ 0.017  & 14.962 $\pm$ 0.025  &  ARIES  \\
2014-03-19.73  &   736.23   &   66.50     & 18.659 $\pm$ 0.156  & 17.306 $\pm$ 0.032  & 16.006 $\pm$ 0.023  & 15.335 $\pm$ 0.017  & 14.980 $\pm$ 0.026  &  ARIES  \\
2014-03-20.71  &   737.21   &   67.48     & 18.931 $\pm$ 0.244  & 17.356 $\pm$ 0.042  & 16.045 $\pm$ 0.027  & 15.371 $\pm$ 0.020  & 15.013 $\pm$ 0.033  &  ARIES  \\
2014-03-21.71  &   738.21   &   68.48     & 18.706 $\pm$ 0.132  & 17.373 $\pm$ 0.032  & 16.046 $\pm$ 0.027  & 15.375 $\pm$ 0.021  & 15.021 $\pm$ 0.034  &  ARIES  \\
2014-03-22.82  &   739.32   &   69.59     & 18.712 $\pm$ 0.173  & 17.406 $\pm$ 0.041  & 16.096 $\pm$ 0.027  & 15.398 $\pm$ 0.020  & 15.046 $\pm$ 0.034  &  ARIES  \\
2014-03-23.82  &   740.32   &   70.59     & 18.825 $\pm$ 0.175  & 17.505 $\pm$ 0.054  &          ---        &          ---        & 15.070 $\pm$ 0.034  &  ARIES  \\
2014-03-29.87  &   746.37   &   76.64     &          ---        &          ---        & 16.350 $\pm$ 0.033  & 15.554 $\pm$ 0.022  & 15.181 $\pm$ 0.035  &  ARIES  \\
2014-04-02.62  &   750.12   &   80.39     &          ---        & 17.814 $\pm$ 0.048  & 16.532 $\pm$ 0.039  & 15.705 $\pm$ 0.025  & 15.291 $\pm$ 0.037  &  ARIES  \\
2014-04-07.76  &   755.26   &   85.53     &          ---        & 18.420 $\pm$ 0.100  & 17.025 $\pm$ 0.057  & 16.107 $\pm$ 0.033  & 15.648 $\pm$ 0.042  &  ARIES  \\
2014-04-08.70  &   756.20   &   86.47     & 19.576 $\pm$ 0.295  & 18.437 $\pm$ 0.075  & 17.222 $\pm$ 0.061  & 16.227 $\pm$ 0.031  & 15.756 $\pm$ 0.035  &  ARIES  \\
2014-04-09.84  &   757.34   &   87.61     & 20.295 $\pm$ 0.582  & 18.610 $\pm$ 0.091  & 17.436 $\pm$ 0.083  & 16.388 $\pm$ 0.041  & 15.902 $\pm$ 0.047  &  ARIES  \\
2014-04-11.76  &   759.26   &   89.53     &          ---        & 18.967 $\pm$ 0.104  & 17.586 $\pm$ 0.084  & 16.587 $\pm$ 0.043  & 16.080 $\pm$ 0.044  &  ARIES  \\
2014-04-13.73  &   761.23   &   91.50     &          ---        &          ---        & 17.863 $\pm$ 0.124  &          ---        &          ---        &  ARIES  \\
2014-04-14.69  &   762.19   &   92.46     &          ---        & 19.200 $\pm$ 0.149  & 17.952 $\pm$ 0.117  & 16.824 $\pm$ 0.052  & 16.306 $\pm$ 0.052  &  ARIES  \\
2014-04-19.72  &   767.22   &   97.49     & 20.076 $\pm$ 0.485  & 19.064 $\pm$ 0.133  & 18.032 $\pm$ 0.141  & 16.943 $\pm$ 0.067  & 16.451 $\pm$ 0.069  &  ARIES  \\
2014-04-20.71  &   768.21   &   98.48     & 19.719 $\pm$ 0.302  & 18.988 $\pm$ 0.121  & 18.125 $\pm$ 0.133  & 16.958 $\pm$ 0.059  & 16.440 $\pm$ 0.057  &  ARIES  \\
2014-04-21.71  &   769.21   &   99.48     & 19.580 $\pm$ 0.260  & 19.074 $\pm$ 0.126  & 18.123 $\pm$ 0.138  & 16.973 $\pm$ 0.060  & 16.482 $\pm$ 0.059  &  ARIES  \\
2014-04-25.80  &   773.30   &  103.57     &          ---        & 19.137 $\pm$ 0.131  & 18.106 $\pm$ 0.156  & 17.007 $\pm$ 0.073  & 16.516 $\pm$ 0.073  &  ARIES  \\
2014-04-26.70  &   774.20   &  104.47     &          ---        & 19.178 $\pm$ 0.128  & 18.148 $\pm$ 0.140  & 17.056 $\pm$ 0.064  & 16.555 $\pm$ 0.063  &  ARIES  \\
2014-04-28.65  &   776.15   &  106.42     & 20.818 $\pm$ 1.294  & 19.074 $\pm$ 0.142  & 18.175 $\pm$ 0.163  & 17.040 $\pm$ 0.077  & 16.597 $\pm$ 0.080  &  ARIES  \\
2014-05-04.69  &   782.19   &  112.46     & 20.558 $\pm$ 1.027  & 19.352 $\pm$ 0.175  & 18.311 $\pm$ 0.180  & 17.121 $\pm$ 0.081  & 16.638 $\pm$ 0.082  &  ARIES  \\
2014-05-09.67  &   787.17   &  117.44     &          ---        & 19.530 $\pm$ 0.176  & 18.429 $\pm$ 0.186  & 17.225 $\pm$ 0.077  & 16.754 $\pm$ 0.078  &  ARIES  \\
2014-05-10.68  &   788.18   &  118.45     &          ---        & 19.738 $\pm$ 0.238  & 18.399 $\pm$ 0.182  & 17.257 $\pm$ 0.080  & 16.796 $\pm$ 0.080  &  ARIES  \\
2014-05-13.67  &   791.17   &  121.44     &          ---        & 19.647 $\pm$ 0.221  & 18.433 $\pm$ 0.213  & 17.292 $\pm$ 0.092  & 16.869 $\pm$ 0.097  &  ARIES  \\
2014-05-15.68  &   793.18   &  123.45     &          ---        & 19.524 $\pm$ 0.215  & 18.516 $\pm$ 0.208  & 17.281 $\pm$ 0.083  & 16.840 $\pm$ 0.086  &  ARIES  \\
2014-05-17.63  &   795.13   &  125.40     &          ---        & 19.404 $\pm$ 0.151  & 18.458 $\pm$ 0.193  & 17.367 $\pm$ 0.100  & 16.910 $\pm$ 0.088  &  ARIES  \\
2014-05-20.67  &   798.17   &  128.44     &          ---        & 19.378 $\pm$ 0.158  & 18.660 $\pm$ 0.228  & 17.390 $\pm$ 0.087  & 17.014 $\pm$ 0.092  &  ARIES  \\
2014-05-21.67  &   799.17   &  129.44     &          ---        & 19.452 $\pm$ 0.163  & 18.693 $\pm$ 0.234  & 17.406 $\pm$ 0.089  & 17.014 $\pm$ 0.093  &  ARIES  \\
2014-05-26.67  &   804.17   &  134.44     &          ---        & 19.516 $\pm$ 0.203  & 18.606 $\pm$ 0.226  & 17.495 $\pm$ 0.117  & 17.122 $\pm$ 0.127  &  ARIES  \\
2014-05-29.68  &   807.18   &  137.45     &          ---        & 19.542 $\pm$ 0.235  & 18.678 $\pm$ 0.272  & 17.509 $\pm$ 0.102  & 17.118 $\pm$ 0.109  &  ARIES  \\
2014-06-15.64  &   824.14   &  154.41     &          ---        &          ---        &          ---        & 17.796 $\pm$ 0.156  & 17.417 $\pm$ 0.167  &  ARIES  \\
2014-06-16.63  &   825.13   &  155.40     &          ---        &          ---        & 18.918 $\pm$ 0.293  & 17.772 $\pm$ 0.148  & 17.430 $\pm$ 0.161  &  ARIES  \\
2014-06-27.65  &   836.15   &  166.42     &          ---        & 20.092 $\pm$ 0.332  & 19.473 $\pm$ 0.501  & 18.014 $\pm$ 0.193  & 17.731 $\pm$ 0.225  &  ARIES  \\
2014-11-26.95  &   988.45   &  318.72     &          ---        &          ---        &          ---        & 19.883 $\pm$ 0.959  & 19.792 $\pm$ 1.411  &  ARIES  \\

\hline

  \end{tabular}
  \\
  \textit{Swift}~UVOT photometry\\
  \setlength{\tabcolsep}{4pt}
  \begin{tabular}
  {c c r c c c c c c c l}
  \hline
    UT Date&JD&Phase$^{a}$     & $uvw2$& $uvm2$& $uvw1$& $uvu$& $uvb$& $uvv$&Tel$^{b}$ & \\
    (yyyy/mm/dd)&2456000+&(day)&(mag)& (mag)& (mag)& (mag)& (mag)& (mag)& /Inst & \\ \hline

2014-01-15.76  &   673.26   &    1.45     & 14.019 $\pm$ 0.048  & 14.007 $\pm$ 0.048  & 13.845 $\pm$ 0.048  & 13.851 $\pm$ 0.044  & 15.032 $\pm$ 0.050  & 15.054 $\pm$ 0.067  &  SWIFT \\
2014-01-16.79  &   674.29   &    2.47     & 14.084 $\pm$ 0.048  &          ---        & 13.803 $\pm$ 0.044  &          ---        &          ---        &          ---        &  SWIFT \\
2014-01-17.19  &   674.69   &    2.87     & 14.115 $\pm$ 0.049  & 14.060 $\pm$ 0.054  & 13.789 $\pm$ 0.046  & 13.651 $\pm$ 0.043  & 14.786 $\pm$ 0.047  & 14.756 $\pm$ 0.068  &  SWIFT \\
2014-01-18.88  &   676.38   &    4.56     & 14.535 $\pm$ 0.054  & 14.326 $\pm$ 0.056  & 13.975 $\pm$ 0.048  & 13.655 $\pm$ 0.043  & 14.799 $\pm$ 0.047  & 14.568 $\pm$ 0.061  &  SWIFT \\
2014-01-19.21  &   676.71   &    4.89     & 14.610 $\pm$ 0.054  & 14.368 $\pm$ 0.056  & 13.987 $\pm$ 0.047  & 13.715 $\pm$ 0.043  & 14.765 $\pm$ 0.047  & 14.729 $\pm$ 0.061  &  SWIFT \\
2014-01-19.47  &   676.97   &    5.15     & 14.662 $\pm$ 0.053  &          ---        &          ---        & 13.725 $\pm$ 0.042  &          ---        &          ---        &  SWIFT \\
2014-01-20.80  &   678.30   &    6.48     & 14.954 $\pm$ 0.061  & 14.768 $\pm$ 0.064  & 14.223 $\pm$ 0.050  & 13.779 $\pm$ 0.043  & 14.767 $\pm$ 0.047  & 14.535 $\pm$ 0.059  &  SWIFT \\
2014-01-21.91  &   679.41   &    7.59     & 15.193 $\pm$ 0.069  & 15.038 $\pm$ 0.064  & 14.347 $\pm$ 0.052  & 13.850 $\pm$ 0.044  & 14.757 $\pm$ 0.047  & 14.601 $\pm$ 0.060  &  SWIFT \\
2014-01-22.96  &   680.46   &    8.65     & 15.474 $\pm$ 0.081  & 15.172 $\pm$ 0.069  & 14.675 $\pm$ 0.067  & 13.858 $\pm$ 0.045  & 14.692 $\pm$ 0.048  & 14.332 $\pm$ 0.058  &  SWIFT \\
2014-01-23.70  &   681.20   &    9.38     & 15.658 $\pm$ 0.087  &          ---        &          ---        & 13.949 $\pm$ 0.043  &          ---        &          ---        &  SWIFT \\
2014-01-23.97  &   681.47   &    9.65     & 15.630 $\pm$ 0.088  & 15.463 $\pm$ 0.080  & 14.780 $\pm$ 0.073  & 13.969 $\pm$ 0.047  & 14.783 $\pm$ 0.050  & 14.356 $\pm$ 0.062  &  SWIFT \\
2014-01-24.87  &   682.37   &   10.55     & 15.829 $\pm$ 0.078  & 15.646 $\pm$ 0.066  & 14.855 $\pm$ 0.063  & 14.004 $\pm$ 0.044  & 14.786 $\pm$ 0.047  & 14.581 $\pm$ 0.058  &  SWIFT \\
2014-01-29.41  &   686.91   &   15.09     & 16.735 $\pm$ 0.087  & 16.591 $\pm$ 0.081  & 15.581 $\pm$ 0.066  & 14.470 $\pm$ 0.048  & 14.975 $\pm$ 0.048  & 14.596 $\pm$ 0.057  &  SWIFT \\
2014-02-07.85  &   696.35   &   24.53     & 18.389 $\pm$ 0.158  & 18.446 $\pm$ 0.157  & 17.035 $\pm$ 0.087  & 15.620 $\pm$ 0.068  & 15.520 $\pm$ 0.056  & 14.963 $\pm$ 0.059  &  SWIFT \\
2014-02-10.18  &   698.68   &   26.86     & 18.636 $\pm$ 0.175  & 18.946 $\pm$ 0.208  & 17.349 $\pm$ 0.097  & 15.884 $\pm$ 0.069  & 15.763 $\pm$ 0.060  & 15.030 $\pm$ 0.059  &  SWIFT \\

  \hline
  \end{tabular}
\begin{flushleft}
  $^{a}$ with reference to the explosion epoch JD=\tg d\\
  $^{b}$ ARIES: 104cm Sampurnanand telescope and 130cm Devasthal fast optical telescope at ARIES, Nainital, Inidia; SWIFT: \textit{Swift}~UVOT\\
  Note: Data observed within 5 Hrs, are represented under single epoch observation.
\end{flushleft}
\end{table*}

%% file: polsn.tex
\begin{table*}
\centering
  \caption{Polarimetric evolution of \sne.}
  \label{tab:pol.SNe}
  \begin{tabular}
  {c c c c c c c c}
  \hline
  UT Date&JD&Phase$^{a}$     & \multicolumn{2}{c}{Observed}    & \multicolumn{2}{c}{$\rm ISP_{MW}$ subtracted}      \\
  (yyyy-mm-dd)&2456000+&(day)&$P_{R}$ (\%)&$\theta_{R}$ (\degr)&$P_{R}$ (\%)&$\theta_{R}$ (\degr)  \\
  \hline
  \multicolumn{7}{c}{\snhj}\\
  \hline
2014-01-02.96  & 660.46 & 23.46 & 0.56 $\pm$ 0.42 & 132.3  $\pm$ 21.8 & 0.63 $\pm$ 0.42 & 127.3  $\pm$ 21.8   \\
2014-01-07.83  & 665.33 & 28.33 & 0.67 $\pm$ 0.50 & 127.1  $\pm$ 18.6 & 0.75 $\pm$ 0.50 & 123.4  $\pm$ 18.6   \\
2014-01-24.85  & 682.35 & 45.35 & 0.44 $\pm$ 0.21 & 150.8  $\pm$ 10.4 & 0.44 $\pm$ 0.21 & 142.5  $\pm$ 10.4   \\
2014-02-05.73  & 694.23 & 57.23 & 0.88 $\pm$ 0.29 & 151.2  $\pm$ ~9.6 & 0.87 $\pm$ 0.29 & 147.0  $\pm$ ~9.6   \\
2014-03-06.69  & 723.19 & 86.19 & 0.98 $\pm$ 0.43 & 128.8  $\pm$ 12.8 & 1.07 $\pm$ 0.43 & 126.1  $\pm$ 12.8   \\
  \hline
  \multicolumn{7}{c}{\sng}\\
  \hline
2014-01-23.92  & 681.42  & 11.72 &   0.82 $\pm$  0.65 &  131.9 $\pm$ 24.7 & 0.78 $\pm$ 0.65 & 134.0  $\pm$ 24.7   \\
2014-01-26.93  & 684.43  & 14.73 &   1.59 $\pm$  0.21 &  132.2 $\pm$ ~4.2 & 1.54 $\pm$ 0.21 & 133.3  $\pm$ ~4.2   \\
2014-02-05.82  & 694.32  & 24.62 &   0.86 $\pm$  0.20 &  134.3 $\pm$ ~6.8 & 0.82 $\pm$ 0.20 & 136.4  $\pm$ ~6.8   \\

\hline

  \end{tabular}

\begin{flushleft}
  $^{a}$ with reference to the explosion epochs JD \thj d and  \tg d for \sne\ respectively.
\end{flushleft}
\end{table*}

%% file: polstar.tex
\begin{table*}
 \centering
  \caption{Polarization measurements in \textit{R}-band for field stars towards the direction of \sne.}
  \label{tab:pol.star}

  \begin{tabular}{lcccc}
     \hline
     Star& $\alpha_{\rm J2000}$&       $\delta_{\rm J2000}$& $ P_R $       & $\theta_R$      \\
       ID&          (h m s)&(\degr\,\arcmin\,\arcsec)      &(\%)      &(\degr)     \\
     \hline
     \multicolumn{5}{c}{Stars towards \snhj.}\\
     \hline
HD 80083    & 09:17:36.04 & -07:27:46.44 &  0.12  $\pm$  0.10  &  ~5.3  $\pm$  25.3 \\
HD 79289    & 09:12:47.60 & -16:47:58.57 &  0.16  $\pm$  0.07  &  10.0  $\pm$  15.1 \\
HD 79308    & 09:12:57.26 & -15:25:09.56 &  0.07  $\pm$  0.09  &  36.4  $\pm$  40.1 \\
HD 82734    & 09:33:12.46 & -21:06:56.60 &  0.25  $\pm$  0.07  &  18.8  $\pm$  8.0~ \\
HD 80749    & 09:21:09.59 & -15:31:37.92 &  0.17  $\pm$  0.10  &  15.1  $\pm$  18.7 \\
HD 78891    & 09:10:10.34 & -16:51:45.40 &  0.03  $\pm$  0.05  &  18.7  $\pm$  42.9 \\
HD 79914    & 09:16:27.54 & -13:50:02.81 &  0.08  $\pm$  0.06  &  17.6  $\pm$  21.3 \\
HD 78920    & 09:10:24.80 & -14:54:00.65 &  0.05  $\pm$  0.05  &  12.1  $\pm$  35.3 \\
HD 77935    & 09:05:02.23 & -11:23:15.73 &  0.15  $\pm$  0.09  &  19.1  $\pm$  18.1 \\
HD 78954    & 09:10:35.01 & -16:58:20.89 &  0.05  $\pm$  0.06  &  27.1  $\pm$  30.7 \\
HD 80990    & 09:22:34.28 & -16:54:15.12 &  0.19  $\pm$  0.09  &  16.3  $\pm$  13.6 \\
     \hline
     \multicolumn{5}{c}{Stars towards \sng.}\\
     \hline
HD 91480    & 10:35:09.69 & +57:04:57.49 &  0.32  $\pm$  0.08  &  17.7  $\pm$  8.7 \\
HD 94247    & 10:53:34.45 & +54:35:06.46 &  0.11  $\pm$  0.06  &  117.1 $\pm$  17.1 \\
HD 102328   & 11:46:55.62 & +55:37:41.48 &  0.22  $\pm$  0.07  &  95.4  $\pm$  8.9 \\

\hline
\end{tabular}

\end{table*}

%% file: ms.bbl
\begin{thebibliography}{94}
\providecommand{\natexlab}[1]{#1}

\bibitem[{{Alard}(2000)}]{2000A&AS..144..363A}
{Alard} C., 2000, \aaps, 144, 363

\bibitem[{{Alard} \& {Lupton}(1998)}]{1998ApJ...503..325A}
{Alard} C., {Lupton} R.~H., 1998, \apj, 503, 325

\bibitem[{{Anderson} et~al.(2014)}]{2014ApJ...786...67A}
{Anderson} J.~P. et~al., 2014, \apj, 786, 67

\bibitem[{{Antezana} et~al.(2013)}]{2013CBET.3757....1A}
{Antezana} R. et~al., 2013, Central Bureau Electronic Telegrams, 3757, 1

\bibitem[{{Arnett}(1980)}]{1980ApJ...237..541A}
{Arnett} W.~D., 1980, \apj, 237, 541

\bibitem[{{Arnett}(1982)}]{1982ApJ...253..785A}
{Arnett} W.~D., 1982, \apj, 253, 785

\bibitem[{{Arnett} \& {Fu}(1989)}]{1989ApJ...340..396A}
{Arnett} W.~D., {Fu} A., 1989, \apj, 340, 396

\bibitem[{{Barbon} et~al.(1979){Barbon}, {Ciatti} \&
  {Rosino}}]{1979A&A....72..287B}
{Barbon} R., {Ciatti} F., {Rosino} L., 1979, \aap, 72, 287

\bibitem[{{Barbon} et~al.(1982){Barbon}, {Ciatti} \&
  {Rosino}}]{1982AA...116...35B}
{Barbon} R., {Ciatti} F., {Rosino} L., 1982, \aap, 116, 35

\bibitem[{{Barbon} et~al.(1990){Barbon}, {Benetti}, {Rosino}, {Cappellaro} \&
  {Turatto}}]{1990A&A...237...79B}
{Barbon} R., {Benetti} S., {Rosino} L., {Cappellaro} E., {Turatto} M., 1990,
  \aap, 237, 79

\bibitem[{{Barrett}(1988)}]{1988MNRAS.234..937B}
{Barrett} P., 1988, \mnras, 234, 937

\bibitem[{{Bayless} et~al.(2013)}]{2013ApJ...764L..13B}
{Bayless} A.~J. et~al., 2013, \apjl, 764, L13

\bibitem[{{Bersten} et~al.(2011){Bersten}, {Benvenuto} \&
  {Hamuy}}]{2011ApJ...729...61B}
{Bersten} M.~C., {Benvenuto} O., {Hamuy} M., 2011, \apj, 729, 61

\bibitem[{{Blondin} \& {Tonry}(2007)}]{2007ApJ...666.1024B}
{Blondin} S., {Tonry} J.~L., 2007, \apj, 666, 1024

\bibitem[{{Bose} \& {Kumar}(2014)}]{2014ApJ...782...98B}
{Bose} S., {Kumar} B., 2014, \apj, 782, 98

\bibitem[{{Bose} et~al.(2013)}]{2013MNRAS.433.1871B}
{Bose} S. et~al., 2013, \mnras, 433, 1871

\bibitem[{{Bose} et~al.(2015{\natexlab{a}})}]{2015MNRAS.450.2373B}
{Bose} S. et~al., 2015{\natexlab{a}}, \mnras, 450, 2373

\bibitem[{{Bose} et~al.(2015{\natexlab{b}})}]{2015ApJ...806..160B}
{Bose} S. et~al., 2015{\natexlab{b}}, \apj, 806, 160

\bibitem[{{Breeveld} et~al.(2011){Breeveld}, {Landsman}, {Holland}, {Roming},
  {Kuin} \& {Page}}]{2011AIPC.1358..373B}
{Breeveld} A.~A., {Landsman} W., {Holland} S.~T., {Roming} P., {Kuin} N.~P.~M.,
  {Page} M.~J., 2011, in J.E. {McEnery}, J.L. {Racusin}, N.~{Gehrels}, eds,
  American Institute of Physics Conference Series. American Institute of
  Physics Conference Series, Vol. 1358, pp. 373--376

\bibitem[{{Brown} et~al.(2014){Brown}, {Breeveld}, {Holland}, {Kuin} \&
  {Pritchard}}]{2014Ap&SS.354...89B}
{Brown} P.~J., {Breeveld} A.~A., {Holland} S., {Kuin} P., {Pritchard} T., 2014,
  \apss, 354, 89

\bibitem[{{Brown} et~al.(2009)}]{2009AJ....137.4517B}
{Brown} P.~J. et~al., 2009, \aj, 137, 4517

\bibitem[{{Burrows}(2013)}]{2013RvMP...85..245B}
{Burrows} A., 2013, Reviews of Modern Physics, 85, 245

\bibitem[{{Chatzopoulos} et~al.(2012){Chatzopoulos}, {Wheeler} \&
  {Vinko}}]{2012ApJ...746..121C}
{Chatzopoulos} E., {Wheeler} J.~C., {Vinko} J., 2012, \apj, 746, 121

\bibitem[{{Chugai}(2006)}]{2006AstL...32..739C}
{Chugai} N.~N., 2006, Astronomy Letters, 32, 739

\bibitem[{{Clocchiatti} \& {Wheeler}(1997)}]{1997ApJ...491..375C}
{Clocchiatti} A., {Wheeler} J.~C., 1997, \apj, 491, 375

\bibitem[{{Denisenko} et~al.(2014)}]{2014CBET.3787....2D}
{Denisenko} D. et~al., 2014, Central Bureau Electronic Telegrams, 3787, 2

\bibitem[{{Dessart} et~al.(2008)}]{2008ApJ...675..644D}
{Dessart} L. et~al., 2008, \apj, 675, 644

\bibitem[{{Eenmae} et~al.(2014){Eenmae}, {Martin}, {Grammer} \&
  {Humphreys}}]{2014ATel.5935....1E}
{Eenmae} T., {Martin} J.~C., {Grammer} S., {Humphreys} R., 2014, The
  Astronomer's Telegram, 5935, 1

\bibitem[{{Elmhamdi} et~al.(2003)}]{2003MNRAS.338..939E}
{Elmhamdi} A. et~al., 2003, \mnras, 338, 939

\bibitem[{{Eswaraiah} et~al.(2013){Eswaraiah}, {Maheswar}, {Pandey}, {Jose},
  {Ramaprakash} \& {Bhatt}}]{2013A&A...556A..65E}
{Eswaraiah} C., {Maheswar} G., {Pandey} A.~K., {Jose} J., {Ramaprakash} A.~N.,
  {Bhatt} H.~C., 2013, \aap, 556, A65

\bibitem[{{Falk} \& {Arnett}(1977)}]{1977ApJS...33..515F}
{Falk} S.~W., {Arnett} W.~D., 1977, \apjs, 33, 515

\bibitem[{{Faran} et~al.(2014)}]{2014MNRAS.445..554F}
{Faran} T. et~al., 2014, \mnras, 445, 554

\bibitem[{{Filippenko}(1997)}]{1997ARA&A..35..309F}
{Filippenko} A.~V., 1997, \araa, 35, 309

\bibitem[{{Filippenko} \& {Foley}(2005)}]{2005IAUC.8484....2F}
{Filippenko} A.~V., {Foley} R.~J., 2005, \iaucirc, 8484, 2

\bibitem[{{Gehrels} et~al.(2004)}]{2004ApJ...611.1005G}
{Gehrels} N. et~al., 2004, \apj, 611, 1005

\bibitem[{{Hamuy}(2003)}]{2003ApJ...582..905H}
{Hamuy} M., 2003, \apj, 582, 905

\bibitem[{{Hamuy} \& {Pinto}(2002)}]{2002ApJ...566L..63H}
{Hamuy} M., {Pinto} P.~A., 2002, \apjl, 566, L63

\bibitem[{{Hamuy} \& {Suntzeff}(1990)}]{1990AJ.....99.1146H}
{Hamuy} M., {Suntzeff} N.~B., 1990, \aj, 99, 1146

\bibitem[{{Han}(2009)}]{2009IAUS..259..455H}
{Han} J., 2009, in K.G. {Strassmeier}, A.G. {Kosovichev}, J.E. {Beckman}, eds,
  IAU Symposium. IAU Symposium, Vol. 259, pp. 455--466

\bibitem[{{Harutyunyan} et~al.(2008)}]{2008A&A...488..383H}
{Harutyunyan} A.~H. et~al., 2008, \aap, 488, 383

\bibitem[{{Heger} et~al.(2003){Heger}, {Fryer}, {Woosley}, {Langer} \&
  {Hartmann}}]{2003ApJ...591..288H}
{Heger} A., {Fryer} C.~L., {Woosley} S.~E., {Langer} N., {Hartmann} D.~H.,
  2003, \apj, 591, 288

\bibitem[{{Heiles}(2000)}]{2000AJ....119..923H}
{Heiles} C., 2000, \aj, 119, 923

\bibitem[{{Hough} et~al.(1987){Hough}, {Bailey}, {Rouse} \&
  {Whittet}}]{1987MNRAS.227P...1H}
{Hough} J.~H., {Bailey} J.~A., {Rouse} M.~F., {Whittet} D.~C.~B., 1987, \mnras,
  227, 1P

\bibitem[{{Inserra} et~al.(2012)}]{2012MNRAS.422.1122I}
{Inserra} C. et~al., 2012, \mnras, 422, 1122

\bibitem[{{Jones} et~al.(2009)}]{2009ApJ...696.1176J}
{Jones} M.~I. et~al., 2009, \apj, 696, 1176

\bibitem[{{Kirshner} \& {Kwan}(1974)}]{1974ApJ...193...27K}
{Kirshner} R.~P., {Kwan} J., 1974, \apj, 193, 27

\bibitem[{{Kumar} et~al.(2014){Kumar}, {Pandey}, {Eswaraiah} \&
  {Gorosabel}}]{2014MNRAS.442....2K}
{Kumar} B., {Pandey} S.~B., {Eswaraiah} C., {Gorosabel} J., 2014, \mnras, 442,
  2

\bibitem[{{Landolt}(2009)}]{2009AJ....137.4186L}
{Landolt} A.~U., 2009, \aj, 137, 4186

\bibitem[{{Leonard} \& {Filippenko}(2001)}]{2001PASP..113..920L}
{Leonard} D.~C., {Filippenko} A.~V., 2001, \pasp, 113, 920

\bibitem[{{Leonard} \& {Filippenko}(2005)}]{2005ASPC..342..330L}
{Leonard} D.~C., {Filippenko} A.~V., 2005, in M.~{Turatto}, S.~{Benetti},
  L.~{Zampieri}, W.~{Shea}, eds, 1604-2004: Supernovae as Cosmological
  Lighthouses. Astronomical Society of the Pacific Conference Series, Vol. 342,
  p. 330

\bibitem[{{Leonard} et~al.(2001){Leonard}, {Filippenko}, {Ardila} \&
  {Brotherton}}]{2001ApJ...553..861L}
{Leonard} D.~C., {Filippenko} A.~V., {Ardila} D.~R., {Brotherton} M.~S., 2001,
  \apj, 553, 861

\bibitem[{{Leonard} et~al.(2002{\natexlab{a}})}]{2002AJ....124.2490L}
{Leonard} D.~C. et~al., 2002{\natexlab{a}}, \aj, 124, 2490

\bibitem[{{Leonard} et~al.(2002{\natexlab{b}})}]{2002PASP..114...35L}
{Leonard} D.~C. et~al., 2002{\natexlab{b}}, \pasp, 114, 35

\bibitem[{{Leonard} et~al.(2006)}]{2006Natur.440..505L}
{Leonard} D.~C. et~al., 2006, \nat, 440, 505

\bibitem[{{Li} et~al.(2011)}]{2011MNRAS.412.1441L}
{Li} W. et~al., 2011, \mnras, 412, 1441

\bibitem[{{Litvinova} \& {Nadezhin}(1985)}]{1985SvAL...11..145L}
{Litvinova} I.~Y., {Nadezhin} D.~K., 1985, Soviet Astronomy Letters, 11, 145

\bibitem[{{Makarov} et~al.(2014){Makarov}, {Prugniel}, {Terekhova}, {Courtois}
  \& {Vauglin}}]{2014A&A...570A..13M}
{Makarov} D., {Prugniel} P., {Terekhova} N., {Courtois} H., {Vauglin} I., 2014,
  \aap, 570, A13

\bibitem[{{Maund} et~al.(2007){Maund}, {Wheeler}, {Patat}, {Wang}, {Baade} \&
  {H{\"o}flich}}]{2007ApJ...671.1944M}
{Maund} J.~R., {Wheeler} J.~C., {Patat} F., {Wang} L., {Baade} D.,
  {H{\"o}flich} P.~A., 2007, \apj, 671, 1944

\bibitem[{{Nagy} et~al.(2014){Nagy}, {Ordasi}, {Vink{\'o}} \&
  {Wheeler}}]{2014A&A...571A..77N}
{Nagy} A.~P., {Ordasi} A., {Vink{\'o}} J., {Wheeler} J.~C., 2014, \aap, 571,
  A77

\bibitem[{{Ochner} et~al.(2014)}]{2014ATel.5767....1O}
{Ochner} P. et~al., 2014, The Astronomer's Telegram, 5767, 1

\bibitem[{{Olivares} et~al.(2010)}]{2010ApJ...715..833O}
{Olivares} E.~F. et~al., 2010, \apj, 715, 833

\bibitem[{{Pastorello} et~al.(2009)}]{2009MNRAS.394.2266P}
{Pastorello} A. et~al., 2009, \mnras, 394, 2266

\bibitem[{{Pereyra} et~al.(2006){Pereyra}, {Magalh{\~a}es}, {Rodrigues},
  {Silva}, {Campos}, {Hickel} \& {Cieslinski}}]{2006A&A...454..827P}
{Pereyra} A., {Magalh{\~a}es} A.~M., {Rodrigues} C.~V., {Silva} C.~R., {Campos}
  R., {Hickel} G., {Cieslinski} D., 2006, \aap, 454, 827

\bibitem[{{Popov}(1993)}]{1993ApJ...414..712P}
{Popov} D.~V., 1993, \apj, 414, 712

\bibitem[{{Poznanski} et~al.(2012){Poznanski}, {Prochaska} \&
  {Bloom}}]{2012MNRAS.426.1465P}
{Poznanski} D., {Prochaska} J.~X., {Bloom} J.~S., 2012, \mnras, 426, 1465

\bibitem[{{Pumo} \& {Zampieri}(2011)}]{2011ApJ...741...41P}
{Pumo} M.~L., {Zampieri} L., 2011, \apj, 741, 41

\bibitem[{{Quimby} et~al.(2007){Quimby}, {Wheeler}, {H{\"o}flich}, {Akerlof},
  {Brown} \& {Rykoff}}]{2007ApJ...666.1093Q}
{Quimby} R.~M., {Wheeler} J.~C., {H{\"o}flich} P., {Akerlof} C.~W., {Brown}
  P.~J., {Rykoff} E.~S., 2007, \apj, 666, 1093

\bibitem[{{Ramaprakash} et~al.(1998){Ramaprakash}, {Gupta}, {Sen} \&
  {Tandon}}]{1998A&AS..128..369R}
{Ramaprakash} A.~N., {Gupta} R., {Sen} A.~K., {Tandon} S.~N., 1998, \aaps, 128,
  369

\bibitem[{{Rautela} et~al.(2004){Rautela}, {Joshi} \&
  {Pandey}}]{2004BASI...32..159R}
{Rautela} B.~S., {Joshi} G.~C., {Pandey} J.~C., 2004, Bulletin of the
  Astronomical Society of India, 32, 159

\bibitem[{{Roming} et~al.(2005)}]{2005SSRv..120...95R}
{Roming} P.~W.~A. et~al., 2005, \ssr, 120, 95

\bibitem[{{Sahu} et~al.(2006){Sahu}, {Anupama}, {Srividya} \&
  {Muneer}}]{2006MNRAS.372.1315S}
{Sahu} D.~K., {Anupama} G.~C., {Srividya} S., {Muneer} S., 2006, \mnras, 372,
  1315

\bibitem[{{Sanders} et~al.(2015)}]{2015ApJ...799..208S}
{Sanders} N.~E. et~al., 2015, \apj, 799, 208

\bibitem[{{Scarrott} et~al.(1990){Scarrott}, {Rolph} \&
  {Semple}}]{1990IAUS..140..245S}
{Scarrott} S.~M., {Rolph} C.~D., {Semple} D.~P., 1990, in R.~{Beck},
  R.~{Wielebinski}, P.P. {Kronberg}, eds, Galactic and Intergalactic Magnetic
  Fields. IAU Symposium, Vol. 140, pp. 245--251

\bibitem[{{Scarrott} et~al.(1991){Scarrott}, {Rolph}, {Wolstencroft} \&
  {Tadhunter}}]{1991MNRAS.249P..16S}
{Scarrott} S.~M., {Rolph} C.~D., {Wolstencroft} R.~W., {Tadhunter} C.~N., 1991,
  \mnras, 249, 16P

\bibitem[{{Schlafly} \& {Finkbeiner}(2011)}]{2011ApJ...737..103S}
{Schlafly} E.~F., {Finkbeiner} D.~P., 2011, \apj, 737, 103

\bibitem[{{Schmidt} et~al.(1992){Schmidt}, {Elston} \&
  {Lupie}}]{1992AJ....104.1563S}
{Schmidt} G.~D., {Elston} R., {Lupie} O.~L., 1992, \aj, 104, 1563

\bibitem[{{Serkowski} et~al.(1975){Serkowski}, {Mathewson} \&
  {Ford}}]{1975ApJ...196..261S}
{Serkowski} K., {Mathewson} D.~S., {Ford} V.~L., 1975, \apj, 196, 261

\bibitem[{{Smartt}(2009)}]{2009ARA&A..47...63S}
{Smartt} S.~J., 2009, \araa, 47, 63

\bibitem[{{Smartt} et~al.(2009){Smartt}, {Eldridge}, {Crockett} \&
  {Maund}}]{2009MNRAS.395.1409S}
{Smartt} S.~J., {Eldridge} J.~J., {Crockett} R.~M., {Maund} J.~R., 2009,
  \mnras, 395, 1409

\bibitem[{{Tomasella} et~al.(2013)}]{2013MNRAS.434.1636T}
{Tomasella} L. et~al., 2013, \mnras, 434, 1636

\bibitem[{{Trammell} et~al.(1993){Trammell}, {Hines} \&
  {Wheeler}}]{1993ApJ...414L..21T}
{Trammell} S.~R., {Hines} D.~C., {Wheeler} J.~C., 1993, \apjl, 414, L21

\bibitem[{{Tran} et~al.(1997){Tran}, {Filippenko}, {Schmidt}, {Bjorkman},
  {Jannuzi} \& {Smith}}]{1997PASP..109..489T}
{Tran} H.~D., {Filippenko} A.~V., {Schmidt} G.~D., {Bjorkman} K.~S., {Jannuzi}
  B.~T., {Smith} P.~S., 1997, \pasp, 109, 489

\bibitem[{{Tully} \& {Fisher}(1988)}]{1988ang..book.....T}
{Tully} R.~B., {Fisher} J.~R., 1988, {Catalog of Nearby Galaxies}

\bibitem[{{Turatto} et~al.(2003){Turatto}, {Benetti} \&
  {Cappellaro}}]{2003fthp.conf..200T}
{Turatto} M., {Benetti} S., {Cappellaro} E., 2003, in W.~{Hillebrandt},
  B.~{Leibundgut}, eds, From Twilight to Highlight: The Physics of Supernovae.
  p. 200

\bibitem[{{Utrobin}(2007)}]{2007A&A...461..233U}
{Utrobin} V.~P., 2007, \aap, 461, 233

\bibitem[{{Valenti} et~al.(2014)}]{2014MNRAS.438L.101V}
{Valenti} S. et~al., 2014, \mnras, 438, L101

\bibitem[{{Valenti} et~al.(2015)}]{2015MNRAS.448.2608V}
{Valenti} S. et~al., 2015, \mnras, 448, 2608

\bibitem[{{Wang} \& {Wheeler}(1996)}]{1996ApJ...462L..27W}
{Wang} L., {Wheeler} J.~C., 1996, \apjl, 462, L27

\bibitem[{{Wang} \& {Wheeler}(2008)}]{2008ARA&A..46..433W}
{Wang} L., {Wheeler} J.~C., 2008, \araa, 46, 433

\bibitem[{{Wang} et~al.(2001){Wang}, {Howell}, {H{\"o}flich} \&
  {Wheeler}}]{2001ApJ...550.1030W}
{Wang} L., {Howell} D.~A., {H{\"o}flich} P., {Wheeler} J.~C., 2001, \apj, 550,
  1030

\bibitem[{{Wang} et~al.(2002{\natexlab{a}}){Wang}, {Baade}, {H{\"o}flich} \&
  {Wheeler}}]{2002Msngr.109...47W}
{Wang} L., {Baade} D., {H{\"o}flich} P., {Wheeler} J.~C., 2002{\natexlab{a}},
  The Messenger, 109, 47

\bibitem[{{Wang} et~al.(2002{\natexlab{b}})}]{2002ApJ...579..671W}
{Wang} L. et~al., 2002{\natexlab{b}}, \apj, 579, 671

\bibitem[{{Wang} et~al.(2003)}]{2003ApJ...591.1110W}
{Wang} L. et~al., 2003, \apj, 591, 1110

\bibitem[{{Zampieri} et~al.(2003)}]{2003MNRAS.338..711Z}
{Zampieri} L., {Pastorello} A., {Turatto} M., {Cappellaro} E., {Benetti} S.,
  {Altavilla} G., {Mazzali} P., {Hamuy} M., 2003, \mnras, 338, 711

\end{thebibliography}
